\documentclass[a4paper,11pt]{article}
\pdfoutput=1 % if your are submitting a pdflatex (i.e. if you have
\usepackage{jcappub} % for details on the use of the package, please
\usepackage[T1]{fontenc} % if needed
\usepackage[utf8]{inputenc}
\usepackage{caption}
\usepackage{color}
\usepackage{multicol}
\usepackage{float}
\usepackage{subcaption}
\usepackage[normalem]{ulem}
\usepackage{hyperref}
\usepackage{xcolor}
\title{\boldmath 
$R$-modes as a New Probe of Dark Matter in Neutron Stars
}

\author[a,1]{Swarnim Shirke\note{Corresponding author.},}
\author[a]{Suprovo Ghosh,}
\author[a]{Debarati Chatterjee,}
\author[b]{Laura Sagunski}
\author[b]{and Jürgen Schaffner-Bielich}

 \affiliation[a]{Inter-University Centre for Astronomy and Astrophysics, Post Bag 4, Ganeshkhind, Pune University Campus, Pune 411007, India}
 \affiliation[b]{Institute for Theoretical Physics,
Goethe University, Max-von-Laue-Stra{\ss}e 1, 60438 Frankfurt am Main, Germany}
 
\emailAdd{swarnim@iucaa.in}
\emailAdd{suprovo@iucaa.in}
\emailAdd{debarati@iucaa.in}
\emailAdd{sagunski@itp.uni-frankfurt.de}
\emailAdd{schaffner@astro.uni-frankfurt.de}
\abstract{
In this work, we perform the first systematic investigation of effects of the presence of dark matter on $r$-mode oscillations in neutron stars (NSs).
Using a self-interacting dark matter (DM) model based on the neutron decay anomaly and a hadronic model obtained from the posterior distribution of a recent Bayesian analysis, we impose constraints on the DM self-interaction strength using recent multimessenger astrophysical observations. We also put new constraints on the DM fraction for this model of DM. The constrained DM interaction strength is then used to estimate DM self-interaction cross section and shear viscosity resulting from DM, which is found to be several orders of magnitude smaller than shear viscosity due to hadronic matter.
Assuming chemical equilibrium among DM fermions and neutrons, we estimate the bulk viscosity resulting from the dark decay of neutrons considering different scenarios for the temperature dependence of the reaction rate and investigate the effect on the $r$-mode instability window.
We conclude that DM shear and bulk viscosity may significantly modify the $r$-mode instability window compared with the minimal hadronic viscosities, depending on the temperature dependence of the process. We also found that for the window to be compatible with the X-ray and pulsar observational data, the rate for the dark decay process must be fast.
}

\begin{document}
\maketitle
\flushbottom

\section{Introduction}
\label{sec:intro}

\subsection{Neutron stars as gravitational wave sources}
\label{sec:nsgw}

%\dc{Motivation for the model by Theo Motta; sufficient DM in NS; $\Gamma_n$ gives key ingredient for DM viscosity; other models should be considered}\\

%\ls{Went through this section and made edits. The next person can take over.}

The direct detection of gravitational waves (GWs) has revolutionised the field of astronomy, particularly for compact stars. Until recently, neutron stars (NS) were only observed using electromagnetic radiation at multiple wavelengths, from which their astrophysical observables could be indirectly deduced. Gravitational waves, on the other hand, allow us to directly probe their complex interior composition. The detection of GWs from the first binary NS merger event GW170817~\cite{Abbott2017AGW170817, abbott2017BGW170817multi} along with simultaneous observations of its electromagnetic counterparts revealed a wealth of information about such systems~\cite{abbott2018, abbott2019properties}, ushering in a new era of multi-messenger astronomy. In particular, tidal effects on the waveform during the inspiral phase of the merger allowed us to put important constraints on dense matter properties \cite{annala2018, Most2018, ghosh2022multi, ghosh2022multihyperon, shirke2023}.

The motivation for studying the NS interior composition is that its ultradense matter environment allows us to probe physics under extreme conditions.  NSs are perfect astrophysical laboratories to study cold and dense nuclear matter. The interior densities in NSs can surpass several times normal nuclear matter densities encountered in terrestrial (nuclear and heavy-ion) experiments~\cite{Lattimer2021}. Further, unlike systems which have the same number density of neutrons and protons, NSs are highly isospin asymmetric (many more neutrons than protons). Nuclear matter theories calculating the NS equation of state (EOS) therefore need to interpolate to high densities and finite isospin asymmetries, which allows us to probe the behaviour of dense matter beyond our present understanding (see e.g.~\cite{oertel2017}). It is expected that strangeness containing matter, such as hyperons or deconfined quark matter, can exist as stable constituents of the inner NS core~\cite{Lattimer2021}.
% \js{can we please cite a more recent work here?}.\dc{done}

Neutron stars are important GW sources~\cite{Abbott2017AGW170817}, not only in binary but also in isolated systems. When NSs are perturbed, quasi-normal modes may become unstable, leading to copious emission of GWs. Depending on the restoring forces that bring the star back to equilibrium, these modes can be classified as {\it f-,p-,g-}modes, etc. for the fluid modes (similar to asteroseismology), while {\it w-}modes are pure-space time modes (similar to black hole quasi-normal modes). Analogous to oscillation modes observed in the solar or stellar physics, NS asteroseismology allows us to directly probe the NS interior composition.

Particularly interesting are {\it r-}modes, that are generic to all rotating neutron stars and restored by the Coriolis force. They may become unstable~\cite{andersson1998} due to the Chandrasekhar-Friedman-Schutz (CFS) mechanism~\cite{CFS1,CFS2}, where more and more negative angular momentum is removed from the system, resulting in an instability and copious GW emission~\cite{andersson2003}. These continuous GWs may be detected by GW detectors such as future observing runs of the LIGO-Virgo-KAGRA detectors, or third generation detectors such as Einstein Telescope~\cite{puntoro2010ET, hild2011ET} or Cosmic Explorer\cite{reitze2019CE, Evans2021}.
% \ls{I like that! Do we want to add references for ET and CE?}  \sss{added}

However, shear and bulk viscosities inside the NS can lead to the damping of the GW emission~\cite{owen1998gravitational, Lindblom2002, Debi_2006, KaonBV, LindblomOwenMorsink1998}.
\iffalse
if the following condition is satisfied:
\begin{equation}
\frac{1}{\tau_{GW} (\Omega,T)} + \frac{1}{\tau_{SV}(\Omega,T)} + \frac{1}{\tau_{BV}(\Omega,T)} = 0~.
\end{equation}
\fi
In NSs, shear viscosity (SV) can result from scattering among leptons, while bulk viscosity (BV) can result from leptonic weak interaction processes. If strangeness-containing matter (such as hyperons or quarks) are present in the core, non-leptonic weak interactions could give rise to additional contributions to bulk viscosity. At low temperatures, superfluidity in hadronic or quark matter would also modify such weak interaction rates, as these viscosities depend on the NS internal composition, and therefore detected GWs would contain the signatures of such composition. $R$-mode oscillations are interesting as they control the maximum observed rotation frequencies of pulsars and also are sources of continuous gravitational waves (CGW)  targeted by searches by the current network of GW detectors~\cite{fesik2020, fesik2020erratum, abbott2021, rajbhandari2021}.

\subsection{Dark matter in neutron stars}
\label{sec:dmns}

%\ls{Went through this section and made edits. The next person can take over.}

DM is the dominant form of matter in the universe, being about five times as prevalent as ordinary nuclear matter. Despite its ubiquity, the particle nature of DM is still unknown. The evidence for DM ranges from astrophysical observations, e.g., of dwarf galaxies rotation curves or the mass distribution of the Bullet cluster, to cosmological observations, such as structure formation and the cosmic microwave background (see, e.g., \cite{Bertone:2018krk,AlvesBatista:2021eeu, khlopov2013, khlopov2021, zurek2014asymmetric, Tulin2018} for reviews). 

It is conjectured that DM could also exist within and around NSs, accreted into its interior forming a DM core~\cite{ellis2018, bell2020, bell2021, anzuini2021} or as DM halo exceeding its radius~\cite{Sagunski:2017nzb,Huang:2018pbu,nelson2019, ivanytskyi2020, Zhang:2021mks,karkevandi2021, karkevandi2022}. %Alternatively, pure DM stars could accrete baryonic matter forming a NS~\cite{ellis2018}. 
The presence of DM in NSs would affect NS observables such as their mass, radius, tidal deformability, or cooling~\cite{kouvaris2008, li2012, sedrakian2016, rezaei2017, panotopoulos2017, ellis2018, sedrakian2019, gresham2019, hcdas2020, karkevandi2021, karkevandi2022, hcdas2022, garcia2022, leung2022, giangrandi2022, diedrichs2023}.  
%Thus, multi-messenger observations of NSs can lead to important constraints of DM particle mass and couplings~\cite{Baryakhtar:2022hbu}.
%\\
%\dc{Review of DM in NSs to be added:} Overview of DM models (including experimental constraints from astrophysics, particle physics and cosmology): \cite{Bertone:2018krk, AlvesBatista:2021eeu} (In \cite{AlvesBatista:2021eeu}, see Sec.9 9 "Dark Matter".)\\
%Review on SIDM (including constraints from astrophysics and particle physics): \cite{Tulin:2017ara} \\
%Gravitational wave probes of dark matter (including dark matter halos around BHs and NSs, exotic compact objects): \cite{Bertone:2019irm} \\
%DM in the NS core: \cite{ellis2018} \\
%NS surrounded by axion cloud: \cite{Huang:2018pbu,Zhang:2021mks} \\
%LIGO constraints on ultralight bosons as DM candidates: \cite{LIGOScientific:2021jlr} 
%\\
%
If DM is present in NSs, it could affect gravitational wave emission through its effects on the global properties such as NS mass and radius, or its effect on unstable oscillation modes such as $r$-modes. 
NS multi-messenger observations could then be used to constraints the DM particle mass and couplings~\cite{Baryakhtar:2022hbu}.

In recent years, there has been significant interest in this topic, and several investigations tried to impose constraints on DM models using astrophysical data~\cite{garani2019, nelson2019, ivanytskyi2020, husain2022a, rutherford2022, shakeri2022}. Moreover, Atreya~et al.~\cite{atreya2019} investigated the effect of DM shear viscosity on cosmological evolution. However, to our knowledge, there exist no detailed investigations about the effect of DM on NS $r$-modes. There have been only two preliminary studies~\cite{horowitz2012, yoshida2020} relating the DM self-interaction cross section with shear viscosity, but the consequent effect on $r$-mode instability is yet to be worked out in detail.

If DM is present in the NS core, it would affect its global properties such as mass, radius, tidal deformability. By comparing with current astrophysical data, one can then constrain the DM model considered. The presence of DM could also affect the transport properties in the NS, such as shear or bulk viscosities. These play a crucial role in determining the stability of the star against $r$-modes and the resulting GW emission. The aim of this work is to perform the first systematic investigation of the effect of DM on $r$-mode oscillations in neutron stars (NS) and the associated gravitational wave (GW) emission. 

This paper is structured as follows. After having outlined the motivation for our work in Sec.~\ref{sec:intro}, we elaborate the models used to describe the internal composition (hadronic and dark matter) and the calculation of the NS structure parameters in Sec.~\ref{sec:models}. In Sec.~\ref{sec:results}, we present the results of this investigation. Finally, in Sec.~\ref{sec:discussions}, we discuss the implications of this study and future directions.

%%%%%%%%%%%%%%%%%%%%%%%%%%%%%%%%%%%%%%%%%%%%%%%
\section{Neutron Star model}
\label{sec:models}

%\dc{Intro: In this section, we describe the NS model applied in this work; Hadronic, DM model, Global structure}
%\ls{Went through this section and made edits. The next person can take over.}

As described in Sec.~\ref{sec:intro}, in this work, we construct theoretical models of neutron stars that describe both hadronic matter as well as DM. There exist various classes of theoretical hadronic matter models, such as microscopic (ab-initio) or phenomenological ones where the model parameters are fitted to reproduce experimental data. The phenomenological models have been particularly successful in reproducing nuclear and hypernuclear experimental data as well as current multi-messenger astrophysical data. Therefore, in this investigation, we adopt such a phenomenological EOS model, the Relativistic Mean Field (RMF) model (see Sec.~\ref{sec:hadronic} for further details).

Various DM models can be constructed 
%and have been explored
%, e.g. fermionic DM, fermion-boson stars, $\phi^4$ theory, DM interacting via Higgs model, superfluid DM, etc 
to describe DM in NSs~\cite{Goldman:1989nd,kouvaris2008,Kouvaris:2010vv,Sandin:2008db,Ciarcelluti:2010ji,Leung:2011zz,Guver:2012ba,Li:2012ii,Xiang:2013xwa,Tolos:2015qra,Mukhopadhyay:2016dsg,ellis2018,McKeen:2018xwc,Baym:2018ljz,motta2018a, motta2018b,ivanytskyi2020,Bell:2020obw,husain2022a,berryman2022BNVProcesses,Cassing:2022tnn,diedrichs2023, GardnerZakeri2023}. 
In this work, we adopt a DM model based on a recent work by~\cite{motta2018a, motta2018b, husain2022a} which is motivated by the fact that the neutron decay anomaly can be explained via the decay of the neutron to the dark sector~\cite{Fornal2019, fornal2020, FornalGrinstein2018prl}. The motivation for applying such a model is twofold: (i) firstly it can explain sufficient DM fraction in NSs which is an essential requirement for our investigation and (ii) secondly, the decay rate of neutrons fixes a key ingredient for the calculation of viscosity in NSs. However, in future works, studies using other models should be considered. The details of the DM model used in this work is given in Sec.~\ref{sec:DM}.

%%%%%%=======================
\subsection{Hadronic matter model}
\label{sec:hadronic}

%\dc{RMF model, GM1 and Hornick parametrisation: please add}\\ 
%\ls{Went through this section and made edits. The next person can take over.}\\

For the description of hadronic matter, we use the relativistic mean field (RMF) model. In the RMF model, nucleons interact via the exchange of $\sigma$ scalar, $\omega$ vector, and $\mathbf{\rho}$ iso-vector mesons. Within the mean-field approximation, each mesonic field is replaced by its expectation value ($\Bar{\sigma}$, $\Bar{\omega}$, $\Bar{\rho}$). The interaction Lagrangian can be written as
\begin{align}\label{interaction_lagrangian}
\mathcal{L}_{int} &= \sum_{N} \bar\psi_{N}\left[g_{\sigma}\sigma-g_{\omega}\gamma^{\mu}\omega_{\mu}-\frac{g_{\rho}}{2}\gamma^{\mu}\boldsymbol{\tau\cdot\rho}_{\mu}\right]\psi_{N} -\frac{1}{3}bm_{N}(g_{\sigma}\sigma)^{3}-\frac{1}{4}c(g_{\sigma}\sigma)^{4} \nonumber \\ 
&+ \Lambda_{\omega}(g^{2}_{\rho}\boldsymbol{\rho^{\mu}\cdot\rho_{\mu}})(g^{2}_{\omega}\omega^{\nu}\omega_{\nu}) + \frac{\zeta}{4!}(g^{2}_{\omega}\omega^{\mu}\omega_{\mu})^{2}~,
\end{align}
where the summation is over the two nucleons ($N$), i.e., protons ($p$) and neutrons ($n$), $\psi_N$ is the Dirac spinor for the nucleons, $m_{N}$ is the vacuum nucleon mass, $\{\gamma^{i}\}$ are the gamma matrices, $\boldsymbol{\tau}$ are Pauli matrices, and $g_{\sigma}$, $g_{\omega}$, $g_{\rho}$ are meson-nucleon coupling constants. The scalar and vector self-interactions couplings are $b$, $c$, and $\zeta$, respectively, while $\Lambda_{\omega}$ is the coupling for the vector-isovector interaction. $\zeta$ is set to zero in this work as it is known to soften the EOS \cite{mueller1996, tolos2017, pradhan2022zeta} 
% \ls{I don't understand the argument here. Why would it be a problem if the EOS is softened?} \sss{We are using the stiffest EoS in the work anyway, so for that EoS $\zeta$ will be zero as $\zeta$ only softens}. 
The energy density of the nucleonic matter thus obtained is given by \cite{hornick2018}
\begin{align}
\epsilon&=\sum_{N}\frac{1}{8\pi^{2}}\left[k_{F_{N}}E^{3}_{F_{N}}+k^{3}_{F_{N}}E_{F_{N}}
-m^{*4}\ln\left(\frac{k_{F_{N}}+E_{F_{N}}}{m^{*}}\right)\right] +\frac{1}{2}m^{2}_{\sigma}\bar\sigma^{2}+ \frac{1}{2}m^{2}_{\omega}\bar\omega^{2}+\frac{1}{2}m^{2}_{\rho}\bar\rho^{2} \nonumber \\ 
&+ \frac{1}{3}bm_{N}(g_{\sigma}\bar\sigma)^{3} + \frac{1}{4}c(g_{\sigma}\bar\sigma)^{4} + 3\Lambda_{\omega}(g_{\rho}g_{\omega}\bar\rho\bar\omega)^{2} + \frac{\zeta}{8}(g_{\omega}\bar{\omega})^{4}~,  \label{energydensity}
\end{align}
where  $k_{F_{N}}$ is the Fermi momentum, $E_{F_{N}} = \sqrt{k_{F_{N}}^{2} + m^{*2}}$ is the Fermi energy, and $m^{*}=m_{N}-g_{\sigma}\sigma$ is the effective mass. 
% \ls{Explain the index $N$ already in equation~\eqref{interaction_lagrangian}?} \sss{done}. 
The pressure ($P$) is given by the Gibbs-Duhem relation 
\begin{equation} \label{eqn:pressure}
P = \sum_{N}{}\mu_{N}n_{N} - \epsilon~,
\end{equation}
where, $\mu_{N} = E_{F_{N}} + g_{\omega}\bar{\omega} + \frac{g_{\rho}}{2}\tau_{3N}\bar{\rho}$. We add to this the contribution from free electrons and muons. The matter is in beta equilibrium and charge neutral which gives the following conditions on the chemical potentials:
\begin{equation}\label{eqn:betaeqlbm_neutrality}
    \mu_n = \mu_p + \mu_e,~ \mu_\mu = \mu_e,~ n_p = n_e + n_\mu
\end{equation}
The coupling constants  ($g_{\sigma}, g_{\omega}, g_{\rho}$, $b$, $c$, and $\Lambda_{\omega}$) are fixed to reproduce the desired nuclear saturation parameters ($n_{sat}$, $E_{sat}$, $K_{sat}$, $E_{sym}$, $L_{sym}$ and $m^*/m$) extracted  from nuclear experiments. In this investigation, we have used two hadronic parametrizations compatible with recent multi-messenger data (see Table.~\ref{table:parametrizations}), HTZCS following~\cite{hornick2018} and another from the results of a recent Bayesian analysis~\cite{ghosh2022multi}, described in further detail below.

\begin{table}[ht]
\centering
\begin{tabular}{c c c c c c c}
\hline\hline 
Model & $n_{sat}$ & $E_{sat}$ & $K_{sat}$ & $E_{sym}$ & $L_{sym}$ & $m^*/m$ \\ 
&(fm$^{-3})$ & (MeV) & (MeV) & (MeV) & (MeV) &\\ \\
[0.5ex] 
\hline 
HTZCS \cite{hornick2018} & 0.15  & -16.0  & 240 & 31 & 50 & 0.65 \\ [1ex]       % [1ex] adds vertical space
% Stiffest \cite{ghosh2022multi} & 0.144972 & -15.966072 & 238.074142 & 31.080246 & 56.482792& 0.550473 \\
Stiffest \cite{ghosh2022multi} & 0.145 & -15.966 & 238.074 & 31.080 & 56.483& 0.550 \\
\hline \hline
\end{tabular}
\caption{Nuclear saturation parameters used in this work. Meson masses are set to $m_{\sigma}=550$ MeV, $m_{\omega}=783$ MeV, $m_{\rho}=770$ MeV, and the nucleon mass is set to $m_N=939.565$ MeV.}
\label{table:parametrizations} 
\end{table}

In recent publications~\cite{ghosh2022multi,ghosh2022multihyperon}, some of the authors of this paper (S. G., D. C. and J.S.B.) explored the parameter space of hadronic matter within the framework of the RMF model  compatible with up-to-date nuclear and hypernuclear experimental data. A hard cut-off scheme with statistical weights was applied to constrain the parameter space imposing multi-physics constraints at different density regimes: chiral effective field theory (CEFT) at low density ($n_b/n_0 \sim 0.4-1$), nuclear and heavy-ion collision data at intermediate density ($n_b/n_0 \sim 1-2$), and multi-messenger astrophysical observations of neutron stars for high density EOS. 
%In a subsequent publication~\cite{shirkeQM2022}, this scheme was extended by the authors (including S.S.) to the quark degrees of freedom incorporating pQCD constraints at very high densities ($n_b/n_0 \geq 40$). \\
%\sg{Add Later based on the graphs in Sec.~\ref{sec:global}.}\\
In this investigation, we have considered the stiffest EOS from the posterior set obtained after imposing the multi-physics constraints (see Table.~\ref{table:parametrizations} "Stiffest" EOS parameter set).

%%%%===================
\subsection{Dark matter model}
\label{sec:DM}
%\dc{Theo Motta's DM model based on neutron decay}\\

The experimental data on the decay of neutrons is not fully understood at present  \cite{czarnecki2018}. The dominant decay channel for neutrons is the $\beta$-decay
$$n \rightarrow p + e^{-} + \bar{\nu_{e}}~.$$
There is a discrepancy in the lifetime of this decay when measured via two different methods: 1) bottle experiments and 2) beam experiments. One possible resolution of the problem is allowing the decay of neutrons into a dark sector \cite{FornalGrinstein2018prl}. 

The Particle Data Group (PDG) quotes the neutron lifetime as  $878.4 \pm 0.5$ s \cite{pdg2022}. It takes into account the recent precise bottle experiment providing an improved measurement of the neutron lifetime as $877.75 \pm 0.28_{\rm stat} +0.22/-0.16_{\rm sys}$ s \cite{gonzalez2021} and seven other previous measurements \cite{serebov2005, pichlmaier2010, steyerl2012, arzumanov2015, serebov2018, pattie2018, ezhov2018}. The beam experiments on the other hand, which count the number of protons emitted, give the neutron lifetime as $888.0 \pm 2.0$ s \cite{czarnecki2018}, based on two beam experiments \cite{byrne1996, Yue2013}. This is about a $4\sigma$ difference in the measurement via  the two methods. It can be explained if the branching fraction of the neutron decay via the $\beta$ decay is less than $100\%$. Fornal $\&$ Grinstein \cite{FornalGrinstein2018prl} suggested that the discrepancy can be resolved if about $1\%$ of neutrons decay into the dark sector. 

In this work, we use a DM model motivated from this anomaly~\cite{FornalGrinstein2018prl}. 
% \ls{Add citations here?} \dc{added same as above} 
One of the mechanisms proposed was a decay involving photon emission $n \rightarrow \chi + \gamma$. Tang~et~al.~\cite{tang2018} performed a follow-up experiment to test this hypothesis and ruled out this possibility of decay. For this study, we use the other decay channel proposed: 
\begin{equation}\label{eqn:darkdecay}
n\rightarrow\chi+\phi~, 
\end{equation}
$\chi$ being the dark spin-half fermion with baryon number 1, and $\phi$ being a light dark boson. This possibility is also explored by various other recent works \cite{motta2018a, motta2018b, husain2022a, Husain2023}. The light dark particle $\phi$ is assumed to escape the NS without any interaction, similar to neutrinos. This sets the equilibrium condition 
\begin{equation}\label{eqn:chemicalequilibrium}
    \mu_{\chi}=\mu_{n}~.
\end{equation}

%\sss{Note that there is another decay channel that was suggested later on by \cite{strumia2022}. Husain~et~al. \cite{husain2022b} focus on this decay channel $n\rightarrow\chi+\chi+\chi$, $\chi$ having baryon number 1/3. The chemical equilibrium in this case satisfies $\mu_{\chi}=\mu_{N}/3$. }

Nuclear stability  requires $937.993$ MeV $< m_{\chi}+m_{\phi}< m_n = 939.565$ MeV \cite{fornal2020}. Stability of the dark particle requires $|m_{\chi} - m_{\phi} | < m_p + m_e = 938.783$ MeV to prevent further beta decay of the dark particles \cite{fornal2020}. We assume $m_{\phi} = 0$ in our model resulting in the condition $937.993$ MeV $< m_{\chi}< 938.783$. We fix the DM mass to $m_{\chi}=938.0$ MeV for this study.

We account for DM self-interactions by adding vector interactions between dark particles, 
\begin{equation}
    \mathcal{L} \supset -g_V\Bar{\chi}\gamma^{\mu}\chi V_{\mu} -\frac{1}{4}V_{\mu\nu}V^{\mu\nu}+ \frac{1}{2}m_V^2V_{\mu}V^{\mu}~,
\end{equation}
where $g_V$ is the coupling strength and $m_V$ is the mass of the vector boson. This results in an additional interaction term in the energy density apart from the free fermion part. The DM energy density is given by
\begin{equation}
    \epsilon_{DM} = \frac{1}{\pi^2}\int_{0}^{k_{F_{\chi}}} k^2\sqrt{k^2 + m_{\chi}^2}dk + \frac{1}{2}Gn_{\chi}^2,
    \label{eqn:endens_dm}
\end{equation}
where,
\begin{equation} \label{eqn:Gdefinition}
    G = \left(\frac{g_V}{m_V}\right)^2, \qquad n_{\chi} = \frac{k_{F_{\chi}}^3}{3 \pi^2}
\end{equation}
The Fermi momentum $k_{F_{\chi}}$ is set by the equilibrium condition Eq.~\ref{eqn:chemicalequilibrium}. Here, $\mu_{\chi} = \sqrt{k_{F_{\chi}}^2+m_{\chi}^2} + Gn_{\chi}$.

% For viscosity calculations we considered the DM .EOS from Motta et al. \cite{motta2018a, motta2018b}
% in which the neutron decay anomaly explained via decay of neutron to dark sector \cite{Fornal2019}.
% This focusses on the decay channel $n\rightarrow\chi+\phi$, $\chi$ being the dark baryon with baryon number 1 and $\phi$ being a light boson. The EOS has been reconstructed using RMF hadronic EOS and solved using the single fluid model unlike the two-fluid approach adopted in Husain et al. \cite{husain2022a}. The reason for not using two-fluid model is that in this case, the DM is considered to be in chemical equilibrium with neutron matter through the condition $\mu_{\chi}=\mu_{N}$. 

% This admixture formalism is similar to previous one apart from the fact that the decay channel is different. This EOS has also been reproduced using a RMF hadronic EOS.

%\dc{Swarnim please add description of RMF EOS in brief and figures for EOS for different g values here}\\
Note that there is another possible decay channel that was suggested later on by \cite{strumia2022} and studied by Husain~et~al. \cite{husain2022b}. This channel involves a decay of the neutron into three DM particles $n\rightarrow\chi+\chi+\chi$, $\chi$ having baryon number 1/3. The chemical equilibrium in this case satisfies $\mu_{\chi}=\mu_{N}/3$. We do not consider this decay channel here.
\\

The light bosons emitted due to the neutron decay (Eq.~\ref{eqn:darkdecay}) carry away momentum and energy, resulting in cooling of the NS. It was shown explicitly in \cite{husain2022a} that the total energy lost as a result of the decay is $\sim$ 0.001 $M_{\odot}$. Assuming that the DM created is roughly degenerate, this excess energy will be carried away by $\phi$, cooling the NS.
% It was explained in \cite{husain2022a} that the total energy lost during this process could result in heating of the DM component by a temperature of few MeV. It was shown that this heating does not change the obsevables significantly. In the extreme case, where the entire energy is taken away by $\phi$ particles, the same amount of energy is lost by the NS.

\begin{figure}[ht]
    \centering
    \includegraphics[scale=0.7]{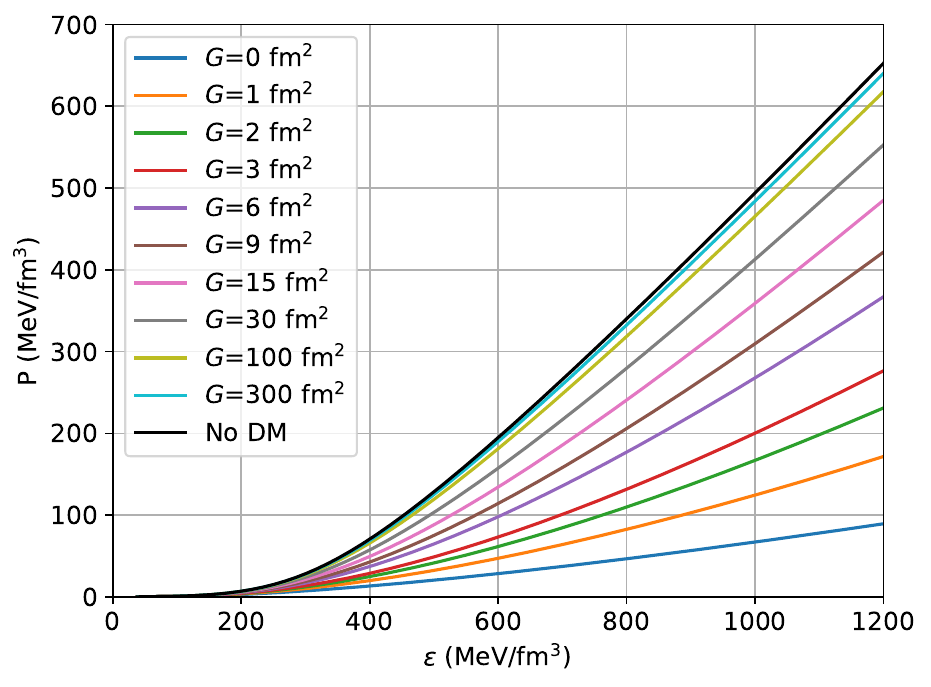}
    \caption{The EOS (pressure vs energy density) for different DM self-interaction strengths $G$. The HTZCS parametrization is used for the nucleonic matter (see Table.~\ref{table:parametrizations}).}
    \label{fig:EOS_hornick}
\end{figure}

\begin{figure}[ht]
    \centering
    \includegraphics[scale=0.7]{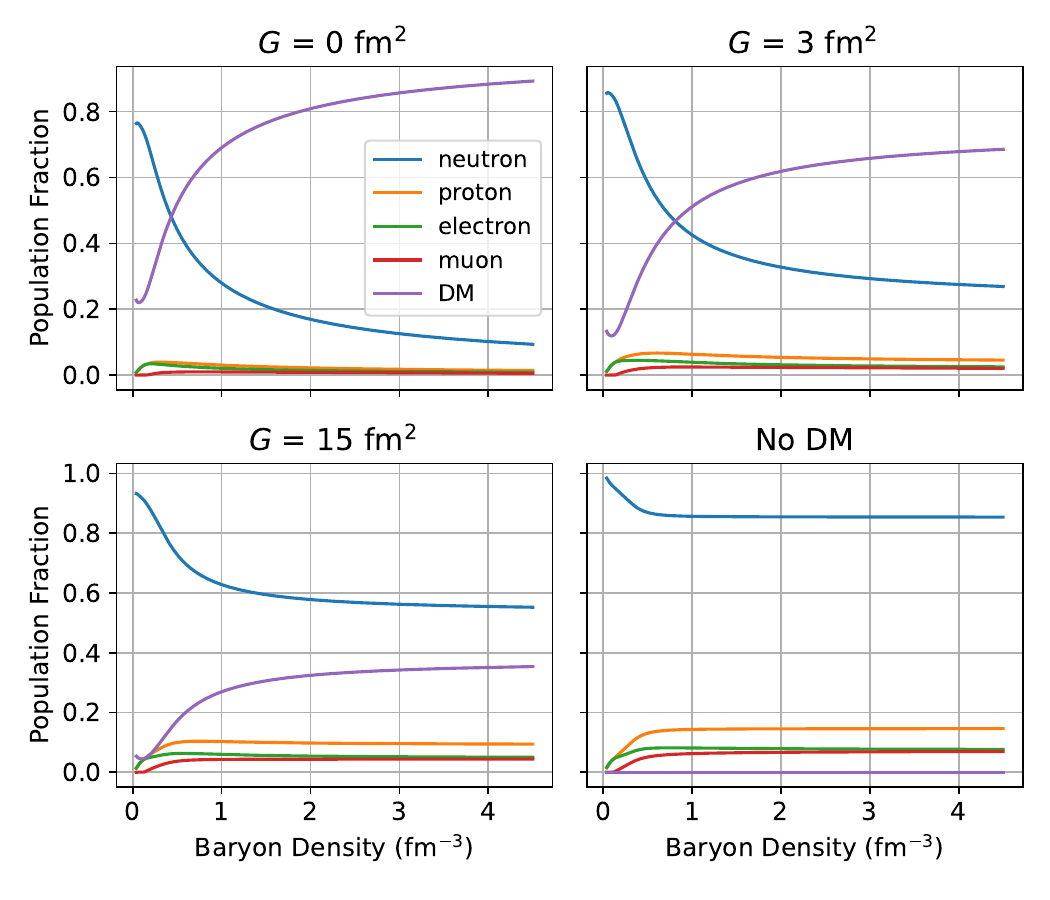}
    \caption{Population fractions of particles for different DM self-interaction strengths $G$. The HTZCS parametrization is used for the nucleonic matter (see Table.~\ref{table:parametrizations}).}
    \label{fig:population_fraction_hornick}
\end{figure}

To check the effect of DM and its self-interaction strength $G$ on the EOS, we plot the EOS for different values of $G$ in Fig. \ref{fig:EOS_hornick}. The presence of DM adds a new degree of freedom and is responsible for softening the EOS. We note that the EOS gets stiffer as the interaction strength is increased, and the DM fraction reduces as we increase $G$. This is evident from the population fractions of various particle species plotted in Fig.~\ref{fig:population_fraction_hornick}. The DM fraction is maximal in the case of non-interacting DM. As we increase $G$, the DM fraction reduces, and for asymptotically large values of $G$, the DM content reduces to zero, that is, the case of pure hadronic EOS without DM. This is consistent with the results from recent studies of DM in NSs \cite{motta2018a, motta2018b}.

\subsection{Global structure}\label{sec:global}
%\dc{Brief discussion on TOV, M-R relation, tidal deformability: please add}\\

To calculate the macroscopic observables like mass and radius, we solve the Tolman–Oppenheimer–Volkoff (TOV) equations
\begin{align}\label{TOV equations}
    \frac{dP(r)}{dr} &= -\frac{[P(r)+\epsilon(r)][m(r)+4\pi r^3P(r)]}{r(r-2m(r))}~, \nonumber \\
    \frac{dm(r)}{dr}&=4\pi r^{2}\epsilon(r)~,
\end{align}
along with the EOS $P(\epsilon)$ described in Sec.~\ref{sec:hadronic} and Sec.~\ref{sec:DM}. The boundary conditions are $m(r=0)=0$ at the centre and $P(r=R)=0$ at the surface. The dimensionless tidal deformability $\Lambda$ is obtained from the expression 
\begin{equation}\label{Lambda}
\Lambda=\frac{2}{3}\frac{k_{2}}{C^{5}}~.
\end{equation}
The $l = 2$ Love number $k_{2}$ is calculated as done in \cite{flanagan2008, hinderer2008, damour2009, yagi2013prd}. Here, we study the case where a chemical equilibrium is established between ordinary matter and the dark sector. So as the density of neutrons drops to zero, the DM population disappears. For this reason, we resort to a single-fluid TOV treatment as opposed to two-fluid as done in \cite{motta2018a, motta2018b, strumia2022}.

%\dc{Swarnim please add figures for M-R and $\Lambda$-M for different g here}\\
In Fig.~\ref{fig:mr_hornick}, we plot the mass-radius curves obtained from solving Eqns.~\ref{TOV equations} for all the EOSs shown in Fig.~\ref{fig:EOS_hornick}. We observe that the maximum mass increases with increasing interaction strength $G$. This variation will be studied in more detail in the next section. We show the bands of the mass measurements for two of the heaviest pulsars observed, PSR J0740+6620 ($M=2.072^{+0.067}_{-0.066}$~\cite{riley2021}) and PSR J0952-0607($M=2.35^{+0.17}_{-0.17}$~\cite{romani2022}), for comparison. The corresponding tidal deformability-mass curves are displayed in Fig.~\ref{fig:lm_hornick} along with the tidal deformability range for 1.4$M_{\odot}$ NS ($\Lambda_{1.4M_{\odot}} = 190^{+390}_{-120}$~\cite{abbott2018}), inferred from the binary neutron star merger event GW170817. Here, we observe that the tidal deformability increases with the interaction strength $G$, too. These results are consistent with previous studies~\cite{motta2018a, husain2022a}.

\iffalse
\begin{figure}[ht]
    \centering
    \includegraphics[scale=0.6]{DM_Rmodes_Figs/mr_hornick_bands.pdf}
    \caption{Mass-Radius curves for different DM self-interaction strengths $G$. The HTZCS parametrization is used for the ordinary hadronic matter (see Table \ref{table:parametrizations})}
    \label{fig:mr_hornick}
\end{figure}

\begin{figure}[ht]
    \centering
    \includegraphics[scale=0.6]{DM_Rmodes_Figs/ml_log_hornick_band.pdf}
    \caption{Tidal deformability ($\Lambda$)--mass curves for different DM self-interaction strengths $G$. The HTZCS parametrization is used for the ordinary hadronic matter (see Table \ref{table:parametrizations})}
    \label{fig:lm_hornick}
\end{figure}
\fi

\begin{figure}[H]
\centering
\begin{minipage}{.49\textwidth}
    \centering
    \includegraphics[width=\textwidth]{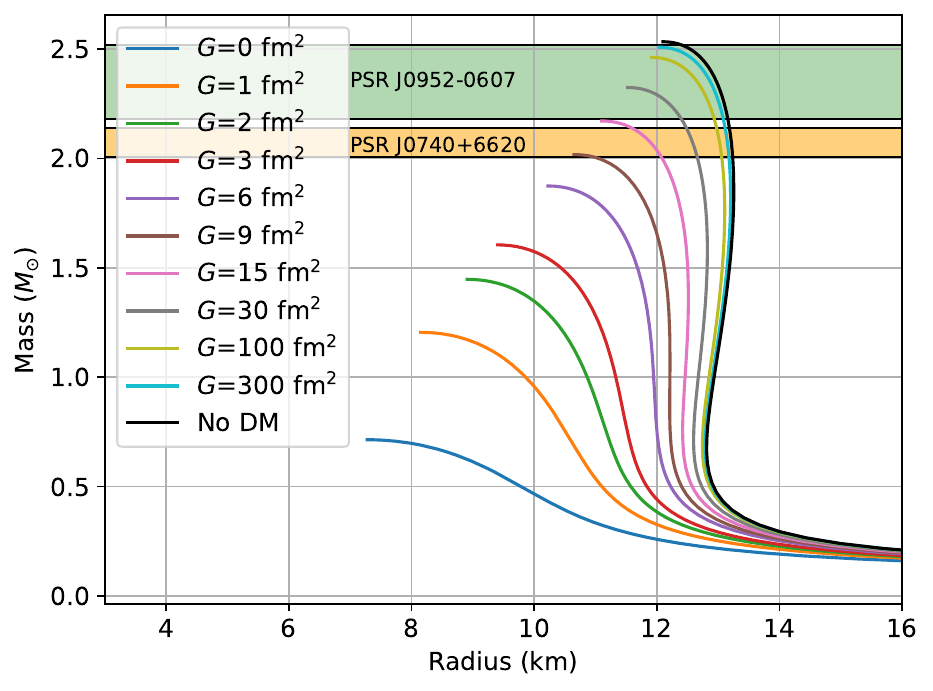}
    \caption{Mass-radius curves for different DM self-interaction strengths $G$. The HTZCS parametrization is used for the  nucleonic matter (see Table \ref{table:parametrizations}). The yellow and green bands correspond to mass measurements of the heaviest pulsars known, $M=2.072^{+0.067}_{-0.066}$ of PSR J0740+6620~\cite{riley2021} and $M=2.35^{+0.17}_{-0.17}$ of PSR J0952-0607~\cite{romani2022} respectively.}
    \label{fig:mr_hornick}
\end{minipage}%
\hfill
\begin{minipage}{.49\textwidth}
    \centering
    \includegraphics[width=\textwidth]{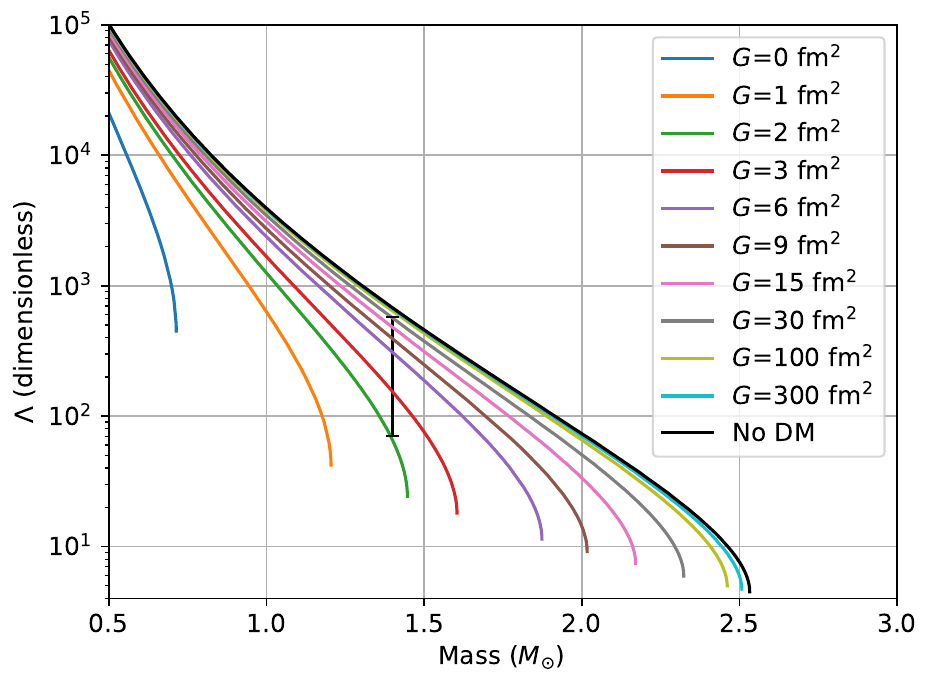}
    \caption{Tidal deformability ($\Lambda$) as a function of mass ($M$) for different DM self-interaction strengths $G$. The HTZCS parametrization is used for the nucleonic matter (see Table \ref{table:parametrizations}). The black bar corresponds to the tidal deformability of a 1.4$M_{\odot}$ NS $\Lambda_{1.4M_{\odot}} = 190^{+390}_{-120}$~\cite{abbott2018}, inferred from the binary neutron star merger event GW170817.}
    \label{fig:lm_hornick}
\end{minipage}
\end{figure}

%%%%%%%%%%%%%%%%%%%%%%%%%%%%%%%%%%%%%%%%%%%%%%%
\section{Results}
\label{sec:results}

\subsection{Constraints on the DM model from NS maximum mass}
\label{sec:constraints}
%\dc{Recap of Bayesian analysis, stiffest EOS satisfying CEFT: Suprovo please add} \\

%\dc{Constraints on g from $M_{max}$ and $\Lambda_{GW170817}$: Swarnim please add}\\

The same DM model considered in this work (see Sec.~\ref{sec:DM}) was also recently applied by Husain et al. \cite{husain2022a}, within the quark-meson coupling (QMC) framework to describe hadronic matter. It was concluded that for a chosen set of coupling parameters, the DM self-interaction strength $G$ should be greater than $26$ fm$^2$ to satisfy the 2-solar-mass constraint for NSs~\cite{riley2021}. 
%\ls{Add reference here?} \dc{done}

In this work, we revisit the analysis using the RMF model for hadronic matter resulting from the Bayesian analysis explained in Sec.~\ref{sec:hadronic}. The resulting nuclear saturation parameters lie within known experimental uncertainties, are consistent with the low-energy CEFT band as well as with multi-messenger astrophysical data. From these parameters, we choose those leading to the stiffest possible hadronic EOS (see `Stiffest' in Table.~\ref{table:parametrizations}). Within this hadronic model, the self-interaction strength $G$ for the DM sector is then varied to produce a set of EOSs with varying stiffness. This enables us to obtain the lowest possible value of $G$ that still satisfies the 2-solar-mass constraint or any maximum mass constraint to be imposed. The mass-radius curves obtained for this parametrization are shown in Fig~\ref{fig:mr_stiffest} for different values of $G$. The maximum mass for this parameter set without any DM content is $\sim 2.94 M_{\odot}$. 
% $G$ is varied on a logarithmic scale. 

Fig.~\ref{fig:maxmass_vs_G} shows how the maximum mass of the DM admixed NS varies with $G$. From this curve, we conclude that for the NS maximum mass to be greater than $2 M_{\odot}$, $G$ has to be greater than 5.6 fm$^2$. If we consider the mass of PSR J0740+6620~\cite{riley2021, fonseca2021}, then $M_{\rm max}\gtrsim2.01 M_{\odot}$ considering the $1\sigma$ interval. This translates to $G\gtrsim5.72$ fm$^2$. For the case of  PSR J0952-0607, the study by \cite{romani2022} impose $M_{\rm max}>2.19 M_{\odot}$ at $1\sigma$ confidence. This gives $G>8.53$ fm$^2$. Thus, the lower limit of $G$ is much lower than $G>26$ fm$^2$ as obtained in \cite{husain2022a}.

\begin{figure}[H]
\centering
% \begin{minipage}{.49\textwidth}
    \centering
    \includegraphics[width=0.7\textwidth]{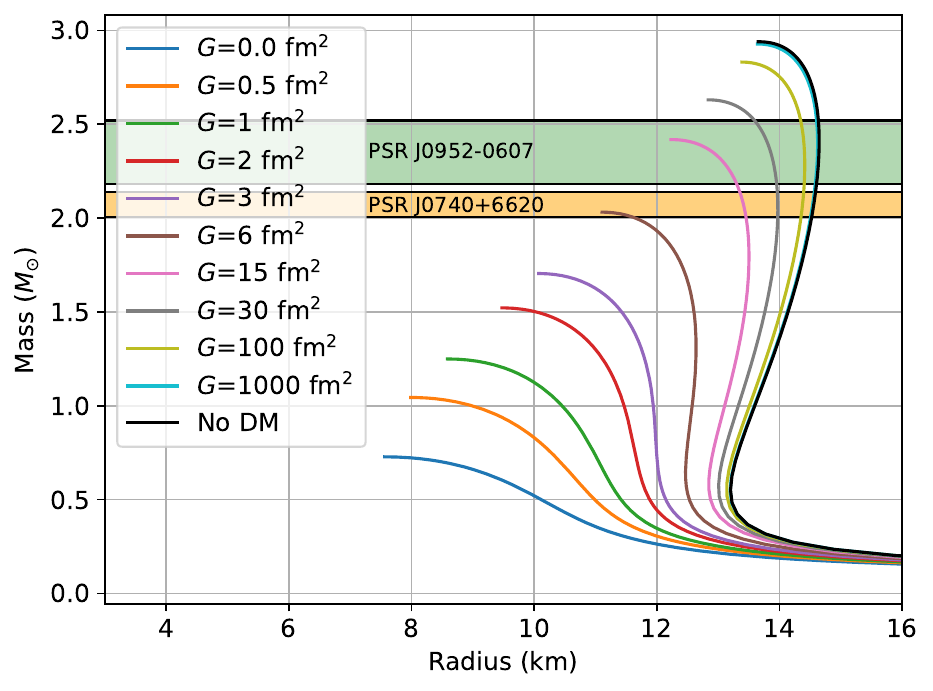}
    \caption{Mass-radius curves for different DM self-interaction strengths $G$. The stiffest parametrization is used for nucleonic matter (see Table. \ref{table:parametrizations}). The yellow and green bands correspond to mass measurements of the heaviest pulsars known, $M=2.072^{+0.067}_{-0.066}$ of PSR J0740+6620~\cite{riley2021} and $M=2.35^{+0.17}_{-0.17}$ of PSR J0952-0607~\cite{romani2022}.}
    \label{fig:mr_stiffest}
\end{figure}

We make a remark on the constraint obtained from tidal deformability of GW170817. From \cite{abbott2018}, we get an upper limit on the tidal deformability of a 1.4$M_{\odot}$ NS ($\Lambda_{1.4M_{\odot}}<580$). As shown in Fig.~\ref{fig:lm_hornick}, $\Lambda$ increases with increasing $G$ and matches that of the pure hadronic case for asymptotically large values of $G$. Thus, if we start with a soft enough hadronic EOS (for which $\Lambda_{1.4M_{\odot}}<580$), the constraint is satisfied by all values of $G$, no matter how large. Thus, we cannot obtain an upper bound on $G$ corresponding to the upper bound on $\Lambda_{1.4M_{\odot}}$. Considering the lower bound on tidal deformability ($\Lambda_{1.4M_{\odot}} = 190^{+390}_{-120}$~\cite{abbott2018}), just as the case for the mass constraint, the stiffest EOS gives the lowest value of $G$ satisfying $\Lambda_{1.4M_{\odot}}>70$. This results in $G>1.6$ fm$^2$. This is lower than $5.6$ fm$^2$ and hence, does not constrain $G$ more than the one from $M_{\rm max}$.
% It is also to be noted that the lower bound of $\Lambda_{1.4M_{\odot}}$ depends on the poorly measured mass ratio of the GW170817 component binaries \cite{kiuchi2019}. This makes this lower bound and the corresponding bound on $G$ unreliable, and we continue using $G>5.6$ fm$^2$ in further discussions.}

\begin{figure}[H]
    \centering
    \includegraphics[width=0.7\textwidth]{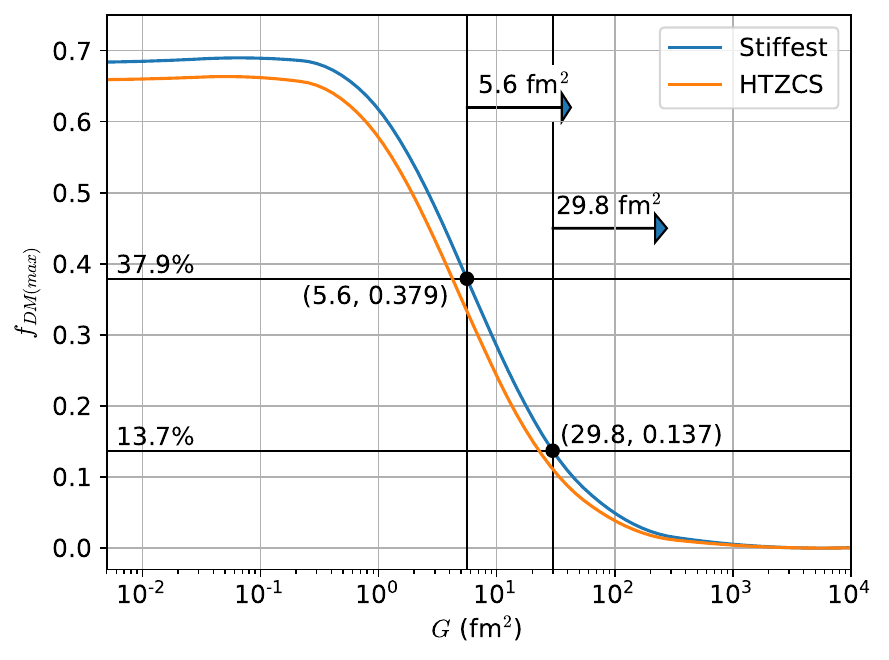}
    \caption{Variation of the maximum DM fraction ($f_{DM(max)}$) with DM self-interaction strength $G$ for the two parameterizations, stiffest and HTZCS (see Table.~\ref{table:parametrizations}). The stiffest parametrization gives the maximum fraction $f_{DM(max)}$  corresponding to the lowest bound on $G$. The fraction $f_{DM(max)}$ for the HTZCS parametrization is also plotted for reference.}
    \label{fig:dmfrac_vs_G}
% \end{minipage}
\end{figure}

From Fig.~\ref{fig:population_fraction_hornick}, it is evident that for a given hadronic EOS, the DM fraction decreases with increase in self-interaction $G$. Corresponding to the ``stiffest'' hadronic EOS, we obtained the lower limit of $G\ge5.6$ fm$^2$ compatible with NS astrophysical data. The obtained constraint on the lower limit of $G$ therefore allows us to comment on the maximum DM fraction ($f_{DM}=M_{\rm DM}/M_{\rm TOT}$) for a DM admixed NS within the neutron decay model of DM. 
%The stiffest EOS (see `Stiffest' in Table.~\ref{table:parametrizations}) results in a maximal DM fraction for a fixed value of $G$. 
%For a fixed EOS parametrization and $G$, the maximum TOV mass configuration results in a DM admixed NS with the highest DM fraction. 
We plot this DM fraction corresponding to the maximum mass configuration ($f_{DM({max)}}$) of the stiffest EOS as a function of $G$ in Fig.~\ref{fig:dmfrac_vs_G}. 
%The DM fraction increases as we decrease $G$. 
%This is consistent with Fig.~\ref{fig:population_fraction_hornick}, where we observe the population fraction to increase with decreasing $G$. 
%The lower bound on $G$ can thus be used to find the maximum possible DM fraction in a DM admixed NS. 
Using $G\ge5.6$ fm$^2$, we get a maximum allowed DM fraction of $37.9\%$. If we consider $G\ge29.8$ fm$^2$, motivated from astrophysical observations (see discussion in Sec.~\ref{sec:crosssection}), we get $f_{DM}\le13.7\%$.

\iffalse
\begin{figure}[ht]
    \centering
    \includegraphics[scale=0.6]{DM_Rmodes_Figs/mr_stiffest_bands.pdf}
    \caption{Mass-radius curves for different DM self-interaction strengths $G$. The stiffest parametrization is used for the ordinary hadronic matter (see Table \ref{table:parametrizations})}
    \label{fig:mr_stiffest}
\end{figure}

\begin{figure}[ht]
    \centering
    \includegraphics[scale=0.6]{DM_Rmodes_Figs/Mmax_vs_G_bands.pdf}
    \caption{Variation of the maximum TOV mass with DM self-interaction strength $G$. The stiffest parametrization is used for the ordinary hadronic matter (see Table \ref{table:parametrizations})}
    \label{fig:maxmass_vs_G}
\end{figure}
\fi

% \end{minipage}%
% \hfill
% \begin{minipage}{.49\textwidth}

\begin{figure}[H]
    \centering
    \includegraphics[width=0.7\textwidth]{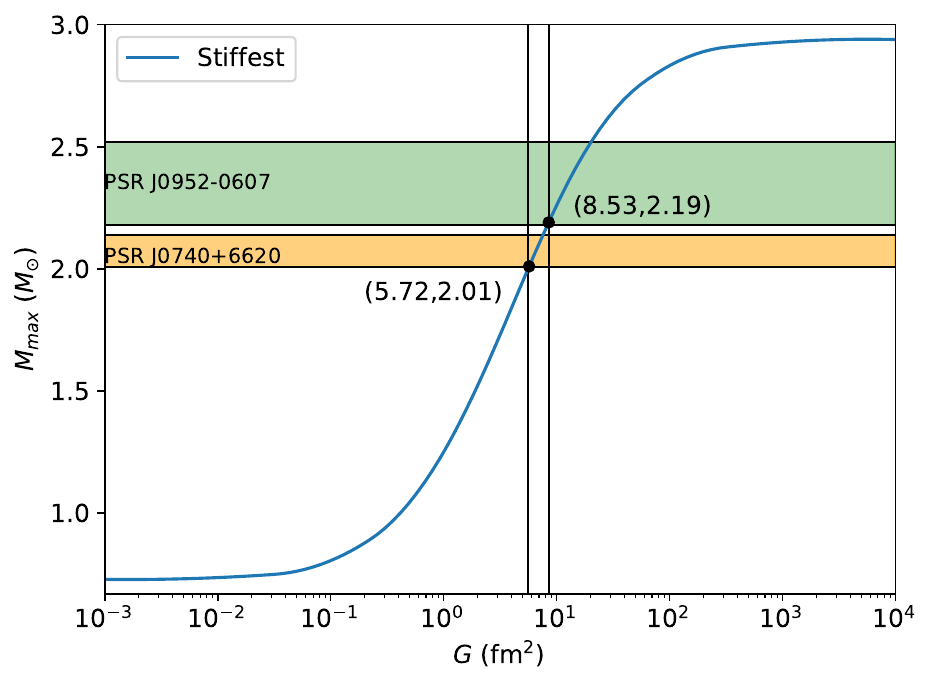}
    \caption{Variation of the maximum TOV mass with DM self-interaction strength $G$. The stiffest parametrization is used for nucleonic matter (see Table. \ref{table:parametrizations}). The yellow and the green bands correspond to mass measurements of the heaviest pulsars known, $M=2.072^{+0.067}_{-0.066}$ of PSR J0740+6620~\cite{riley2021} and $M=2.35^{+0.17}_{-0.17}$ of PSR J0952-0607~\cite{romani2022} respectively.}
    \label{fig:maxmass_vs_G}
% \end{minipage}
\end{figure}

%%%%%%%%%%%%%%%%%%%%%%%%%%%%

\subsection{Shear viscosity in presence of dark matter}
\label{sec:shear}

In non-superfluid nucleonic matter, the primary contribution to shear viscosity (SV) comes from neutron-neutron (n-n) and electron-electron (e-e) scattering~\cite{flowers1976, flowers1979, cutler1987}. To estimate the effect of the presence of DM in NSs on shear viscosity and how it compares with the shear viscosity contribution from n-n and e-e scattering, we employ the kinetic theory formula as done in \cite{horowitz2012, yoshida2020}.
\begin{equation}\label{eqn:shear_formula}
    \eta_{\chi} \approx \frac{\sqrt{m_{\chi}kT}}{\sigma_{\chi}}
\end{equation}
Here, $k$ is the Boltzmann constant and $T$ is temperature. The pre-factor of the expression is  $\mathcal{O}(1)\approx 1$. 
In this study, we consider a self-interaction term of the form $\frac{1}{2}Gn_{\chi}^2$ (see Eq.~\ref{eqn:endens_dm}) which is related to the DM self-interaction cross-section. We calculate this in order to compare it with existing bounds on DM self-interaction cross-section and to get an estimate of shear viscosity. The inter-relation between $G$ and $\sigma_\chi$ is carried out in Sec.\ref{sec:crosssection}.
\\

The shear viscosity given by Eq.\ref{eqn:shear_formula}
involves several assumptions. For the kinetic theory formula to be valid, the shear viscosity must be dominated by the scattering among the dark matter particles as compared to their scattering with neutrons. This is in accordance with calculations of low DM-neutron cross sections~\cite{BertoniNelsonReddy2013, LiuChenJi2017}. It has also been demonstrated that contribution from Pauli blocking and kinematics is non-negligible for low energy and momentum scattering processes~\cite{BertoniNelsonReddy2013}. However, in this work we neglect these effects in order to obtain a first estimate of DM-DM scattering SV. More sophisticated calculations involving collision integrals \cite{flowers1976, Shternin2008} are postponed for future work.

\subsubsection{Dark matter self-interaction cross-section}
\label{sec:crosssection}

Consider self-interacting fermionic DM interacting via a mediator boson $V$ with mass $m_{V}$ and a dark structure constant $\alpha_{\chi}$. Here, $\alpha_{\chi}=g_{V}^2/4\pi$ and $G = g_{V}^2/m_{V}^2$ (see Eq.~\ref{eqn:Gdefinition}). Combining these, we get
\begin{equation}\label{eqn:G_alpha}
    \frac{G}{4\pi}=\frac{\alpha_{\chi}}{m_{V}^2}
\end{equation}
Now from \cite{zurek2014asymmetric, yoshida2020, Tulin2018}, we have the expression for self-interaction cross-section given as

% DM scattering processes can contribute to the shear viscosity in the crust and core ~\cite{horowitz2012,yoshida2020}
% In a preliminary study, Yoshida \cite{yoshida2020} and Narain et al. \cite{Narain:2006kx} used the same self-interacting fermionic DM model to estimate shear viscosity in neutron stars. 
% \\

% For the calculation of shear viscosity, we need the the self-interaction cross-section. 

% \begin{align}\label{eqn:sigmaderivation}
%     \sigma_{\chi} &\approx 5 \times 10^{-23} \left(\frac{\alpha_{\chi}}{0.01}\right)^2\left(\frac{m_{\chi}}{10 \text{ GeV}}\right)^2\left(\frac{10 \text{ MeV}}{m_\phi}\right)^4 \text{cm}^2 \\
%  %    &\approx 5 \times 10^{-17} \left(\frac{\alpha_{\chi} \text{ MeV}^2}{m_{\phi}^2}\right)^2\left(\frac{m_{\chi}}{1 \text{ GeV}}\right)^2 \text{cm}^2 \\
%  %    &\approx 5 \times 10^{-17} \left(\frac{G}{4\pi 
%  % \text{ MeV}^{-2}}\right)^2\left(\frac{m_{\chi}}{1 \text{ GeV}}\right)^2 \text{cm}^2 \\
%     \sigma_{\chi} &\approx 2.1 \times 10^{-28} \left(\frac{G}{1 \text{ fm}^2}\right)^2\left(\frac{m_{\chi}}{1 \text{ GeV}}\right)^2 \text{cm}^2 \\
% \end{align}

\begin{align}\label{eqn:sigmaderivation}
    \sigma_{\chi} &\approx 5 \times 10^{-23} \left(\frac{\alpha_{\chi}}{0.01}\right)^2\left(\frac{m_{\chi}}{10 \text{ GeV}}\right)^2\left(\frac{10 \text{ MeV}}{m_V}\right)^4 \text{cm}^2 \nonumber \\
    %\sigma_{\chi} 
    &\approx 2.1 \times 10^{-28} \left(\frac{G}{1 \text{ fm}^2}\right)^2\left(\frac{m_{\chi}}{1 \text{ GeV}}\right)^2 \text{cm}^2~,
\end{align}
where we make use of Eqn.~\ref{eqn:G_alpha} and $1$ MeV$ = 1/197.3$ fm$^{-1}$. This is valid only when $\alpha_{\chi}m_{\chi}/m_{V} \sim Gm_{\chi}m_{V}\ll 1$ or $m_{V}\ll 1/Gm_{\chi}$. We can write the expression for $\sigma_{\chi}/m_{\chi}$ using the conversion $1$ GeV$ = 1.78 \times 10^{-24}$ g such that
\begin{equation}
    \frac{\sigma_{\chi}}{m_{\chi}} \approx  1.2 \times 10^{-4} \left(\frac{G}{1\text{ fm}^2}\right)^2\left(\frac{m_{\chi}}{1\text{ GeV}}\right)~\text{cm}^2/\text{g}~.
\end{equation}
Thus, the obtained limit of $G \ge 5.6 \text{ fm}^2$ translates to $\sigma_{\chi}/m_{\chi} \ge 3.53 \times 10^{-3}~\text{cm}^2/\text{g}$. Even for $G > 8.53 \text{ fm}^2$ we get $\sigma_{\chi}/m_{\chi} \ge 8.19 \times 10^{-3}~\text{cm}^2/\text{g}$. %This is lower than the lower limit coming from other astrophysical considerations
%\cite{tulin2013, Tulin2018} 
 %\dc{give references?} \sss{Could Laura/Robin/Edwin add these?} \ls{Yes, for sure. I'll do that.} \\ 
%\sss{Thanks. Should we also have a short paragraph on astrophysical constraints on $\sigma/m$ and why we use this range?}\ls{Done. I've added the paragraph below which contains the references and a discussion why we use the particular range for the self-interaction cross section.}
%which give $0.1\le \sigma_{\chi}/m_{\chi} \le 10$ (cm$^2$g$^{-1}$). 

This is lower than the lower limit coming from other astrophysical considerations. Galaxy cluster observations require cross sections of order $\sim 0.1~\text{cm}^2/\text{g}$~\citep{Loeb:2010gj,Kaplinghat:2015aga,Sagunski:2020spe}. For solving the core-cusp problem on galactic scales, on the other hand, cross sections $\lesssim 100~\text{cm}^2/\text{g}$ are needed~\citep{Kaplinghat:2015aga,Sagunski:2020spe}. This is why in this work, we consider the self-interaction cross section range as $0.1~\text{cm}^2/\text{g} \le \sigma_{\chi}/m_{\chi} \le 100~\text{cm}^2/\text{g}$.

% \ls{IMPORTANT: I looked at the literature~\citep{Kaplinghat:2015aga,Sagunski:2020spe} again, and I think, it would be more consistent to use $0.1~\text{cm}^2/\text{g} \le \sigma_{\chi}/m_{\chi} \le 100~\text{cm}^2/\text{g}$ instead of $0.1~\text{cm}^2/\text{g} \le \sigma_{\chi}/m_{\chi} \le 10~\text{cm}^2/\text{g}$ as the cross section range (see also the discussion on p.13 in my paper~\citep{Sagunski:2020spe}). If you agree, we need to update the upper limit on $G$.}
Conversely, these limits on $\sigma_{\chi}/m_{\chi}$ translate to 
% $$28.9 \sqrt{\frac{1\text{GeV}{m_{\chi}}} \text{fm}^2 \le G \le 289 \sqrt{\frac{1\text{GeV}}{m_{\chi}}} \text{fm}^2~.$$
$$28.9 \sqrt{\frac{1\text{GeV}}{m_{\chi}}} \text{fm}^2 \le G \le 914 \sqrt{\frac{1\text{GeV}}{m_{\chi}}} \text{fm}^2~.$$

For DM particle of mass 938 MeV, this becomes
% $$29.8 \text{ fm}^2 \le G \le 298 \text{ fm}^2.$$
$$29.8 \text{ fm}^2 \le G \le 943 \text{ fm}^2.$$

% \sss{If instead, we use the range $0.1~\text{cm}^2/\text{g} \le \sigma_{\chi}/m_{\chi} \le 100~\text{cm}^2/\text{g}$, we get $$28.9 \sqrt{\frac{1GeV}{m_{\chi}}} \text{fm}^2 \le G \le 914 \sqrt{\frac{1GeV}{m_{\chi}}} \text{fm}^2~.$$ And for DM particle of mass 938 MeV, this becomes, $$29.8 \text{ fm}^2 \le G \le 943 \text{ fm}^2.$$}

Thus, the lower limit obtained in Sec.~\ref{sec:constraints} is less constraining than the existing constraints. Using Eqn.~\ref{eqn:shear_formula} and the expression for $\sigma_{\chi}$ derived in Eqn. \ref{eqn:sigmaderivation} we get, 
\begin{align}
    \eta_{\chi} &\approx 5.6\times10^{19}\left(\frac{1\text{fm}^2}{G}\right)\left(\frac{1\text{GeV}}{m_{\chi}}\right)^2\sqrt{m_{\chi}kT} \text{ cm}^{-2} \nonumber \\
    &= 2.6 \times 10^{12} \sqrt{\frac{T}{10^{9}K}}\left(\frac{1\text{fm}^2}{G}\right)^2 \text{ g cm$^{-1}$ s$^{-1}$}~.
\end{align}

\begin{figure}[H]
    \centering
    \includegraphics[scale=0.7]{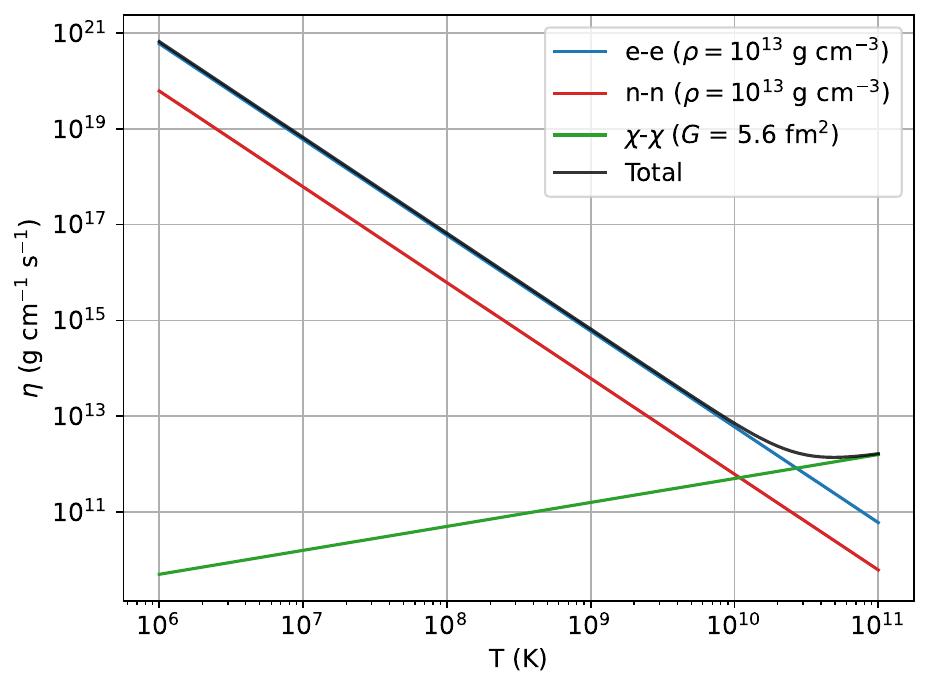}
    \caption{Relative strengths of shear viscosity ($\eta$) from different processes: e-e scattering (blue), n-n scattering (red) and DM-DM scattering (green) as a function of temperature (T).}
    \label{fig:shear}
\end{figure}

The shear viscosity thus obtained, $\eta_{\chi} \sim T^{1/2}G^{-2}$, is orders of magnitude lower than that from n-n scattering ($\eta_{nn} = 347\rho^{9/4}T^{-2}$) and e-e scattering ($\eta_{ee} = 6.0\times10^{6}\rho^{2}T^{-2}$)~\cite{cutler1987}. This is evident from Fig.~\ref{fig:shear}, where we compare the DM shear viscosity with n-n and e-e shear viscosity, at a density $\rho=10^{13}$ g cm$^{-3}$, a value much lower that nuclear saturation density. For the most optimistic case, we use the lowest value of $G$ that was found to be consistent with 2$M_{\odot}$ constraint ($G=5.6$ fm$^2$). Note that this is lower than the lower limit coming from other astrophysical constraints $G = 29.8$ fm$^2$.  For $T<10^9$ K, the DM shear viscosity is more than 2 orders of magnitude lower, and it only becomes important above  $T=10^{10}$ K. At such high temperatures, damping is already dominated by bulk viscosity and is the primary dissipation channel. The actual densities within neutron stars are higher and $G$ is expected to be larger than 5.6 fm$^{2}$, making DM shear viscosity practically negligible. 
% For this reason, we do not consider the contribution of DM SV for the rest of the study.

%%%%%%%%%%%%%%%%%%%%%%%%%%%%%%%%%%%%%%%%%%%%%%%%%%%%%%%%
\subsection{Bulk viscosity in presence of dark matter}
\label{sec:bulk}

%\dc{Suprovo please revise}\\

Bulk viscosity in neutron stars arises from the deviation of the composition from chemical equilibrium due to leptonic or non-leptonic weak interactions. Depending on the timescales of such processes,
bulk viscosity may damp out unstable $r$-modes~\cite{owen1998gravitational}, and the ``instability window" is determined by the balance between gravitational and viscous timescales.
Bulk viscosity of NS matter originating from leptonic processes such as direct URCA~\cite{URCABV1,URCABV2} and modified URCA~\cite{Sawyer1989,mURCABV2} reactions, as well as possible non-leptonic contributions for hyperonic matter~\cite{Jones2001,Lindblom2002,Debi_2006} have been studied extensively. Weak interaction processes involving DM inside the NS core could also potentially contribute to bulk viscosity.  
In our DM model described in Sec.~\ref{sec:DM}, we consider the existence of a dark decay channel including a dark fermion $\chi$ given in Eqn.~\eqref{eqn:darkdecay}, which is very nearly degenerate with the neutron. The presence of this dark fermion $\chi$ changes the composition of the NS as it is in chemical equilibrium with the neutrons.  The corresponding $\beta$-equilibrium equations and charge conservation reactions are given in Eqn.~\eqref{eqn:betaeqlbm_neutrality} and ~\eqref{eqn:chemicalequilibrium}.
For any weak process, the real part of the bulk viscosity coefficient is calculated in terms of relaxation times of microscopic processes
~\cite{Debi_2006}
\begin{equation}\label{BV}
    \zeta = P(\gamma_{\infty} - \gamma_0)\frac{\tau}{1+(\omega\tau)^2}
\end{equation}
where  $P$ stands for the  pressure, $\omega$  for the oscillation frequency of the ($l,m$) r mode which is related to the angular velocity $\Omega$ of a rotating NS~\cite{Provost1981,Ghosh2022,ghosh2023universal}, $\tau$ is the characteristic timescale of the reaction, and $\gamma_{\infty}$ and $\gamma_0$ are the ``infinite" and ``zero" frequency adiabatic indices, respectively. This factor is given by~\cite{Lindblom2002}
\begin{equation}\label{adia}
    \gamma_{\infty} - \gamma_0 = -\frac{n_B}{P}\frac{\partial P}{\partial n_n} \frac{dn_n}{d n_{B}}
\end{equation}
where $n_B$ is the baryon number density and $n_n$ the neutron density. 
\\

Given the EOS, the relaxation timescale ($\tau$)  for the neutron decay~\eqref{eqn:darkdecay} is the only unknown parameter to calculate the bulk viscosity from this neutron decay process. 
The temperature dependence of bulk viscosity depends on the unknown temperature dependence of $\tau$. To investigate the effect of $\tau$ on the bulk viscosity, we first consider two extreme cases, where $\tau$ has the same temperature dependence as m-URCA process (scenario 1) and no temperature dependence (scenario 2) as explained below. The realistic case is expected to lie somewhere between the two extremes. We also consider the intermediate case (scenario 3) where $\tau$ has the same temperature dependence as d-URCA process in NS matter. In general, the matrix elements will also depend on the specific DM model, and the overall equilibration rate becomes arbitrary. Therefore we do not put emphasis on the exact calculation of $\tau$ and leave this for future work.
\\

\begin{figure}[htp] % "[t!]" placement specifier just for this example
\begin{subfigure}{0.48\textwidth}
\includegraphics[width=\linewidth]{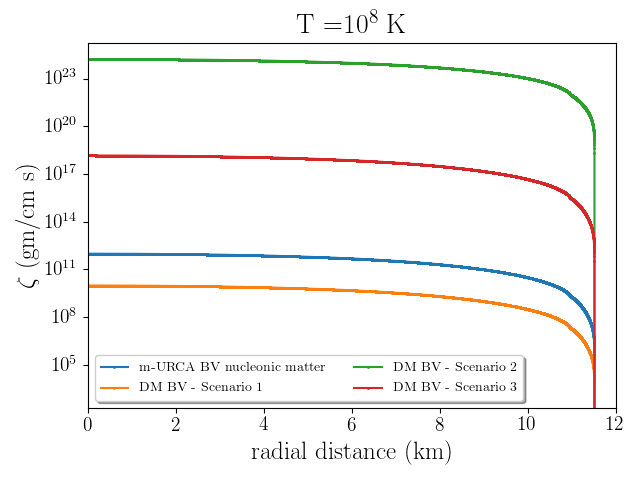}
\end{subfigure}\hspace*{\fill}
\begin{subfigure}{0.48\textwidth}
\includegraphics[width=\linewidth]{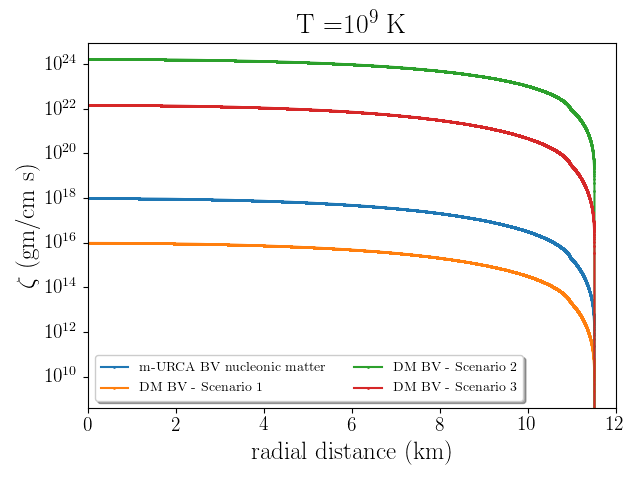}
\end{subfigure}

\medskip
\begin{subfigure}{0.48\textwidth}
\includegraphics[width=\linewidth]{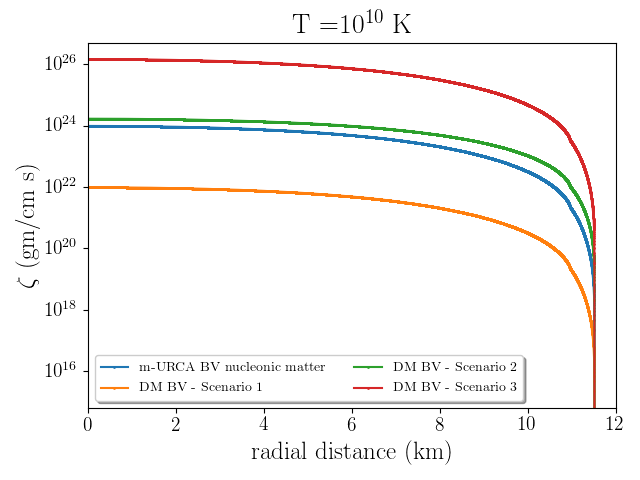}
\end{subfigure}\hspace*{\fill}
\begin{subfigure}{0.48\textwidth}
\includegraphics[width=\linewidth]{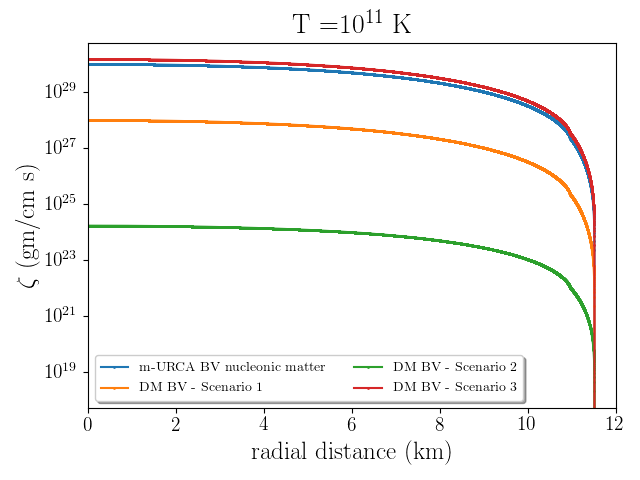}
\end{subfigure}

\caption{Relative strength of the bulk viscosity sources for nucleonic matter and dark matter as function of their radial distances for a 1.4 M$_{\odot}$ NS at different temperatures. For nucleonic matter, only m-URCA bulk viscosity is considered but for DM, three different scenarios (Scenarios - 1, 2 and 3 respectively described in Sec.~\ref{sec:bulk}) of the decay timescale is discussed.}
\label{fig:Strength}
\end{figure}

{\it Scenario 1:} According to the analysis by Fornal and Grinstein et al. (2019)~\cite{Fornal2019}, in order to explain the neutron decay via this particular dark decay channel, the rate for the reaction should be 1/100 times the rate of neutron decay to proton. Assuming that this estimate also remains valid for neutron decay via modified URCA reactions, one can compute the relaxation timescale~\cite{Sawyer1989}  
\begin{equation}\label{tau}
    \frac{1}{\tau_{n\rightarrow \chi}} = \frac{\Delta\Gamma_{n\rightarrow \chi}}{ \delta \mu} \frac{\delta \mu}{n_B\delta x_n} = \frac{\Delta\Gamma_{m-URCA}/100}{ \delta \mu} \frac{\delta \mu}{n_B\delta x_n}~.
\end{equation}
In the calculation of the timescale, we assumed that the relaxation rate is always 100 times smaller than that of the m-URCA reaction, which implies that it follows the same temperature dependence as that of the m-URCA reaction.  The other term $\frac{\delta \mu}{n_B\delta x_n}$ can be directly calculated from the EOS~\cite{Lindblom2002,Debi_2006} numerically. \\
 From the  bulk viscosity calculation for the modified URCA process~\cite{Sawyer1989}, up to leading order in $\delta \mu$, we have
\begin{equation}\label{Sawyer}
    \frac{\Delta\Gamma_{m-URCA}}{\delta \mu} = \lambda_{m-URCA} = 1.62\times 10^{-18}\hbar^{-4}c^3m_e^3\left(\frac{k_BT}{1\text{MeV}}\right)^6~.
\end{equation}
 This calculation shows that, in this scenario, the DM bulk viscosity will always be  several orders of magnitude smaller than that of the one of nucleonic m-URCA  for all temperatures. 
 % \sout{In the left panel of  we compare the strength of the DM bulk viscosity against the m-URCA bulk viscosity at different temperatures. } 
 This is also evident from Fig.~\ref{fig:Strength}, where DM bulk viscosity (orange curve) is compared to that of nucleonic m-URCA BV (blue curve). 
\\

{\it Scenario 2:} The other extreme case is the constant timescale scenario, which can be motivated by the analogy with the non-leptonic weak decay in NSs, $n \to p + K^-$, where $K^-$ is an s-wave condensate \cite{KaonBV}. 
The equilibration rate is given by
\begin{equation}
    \Gamma_{dark} = \int \frac{d^3k_1}{(2\pi)^32E_1}\frac{d^3k_2}{(2\pi)^32E_2}\frac{d^3k_3}{(2\pi)^32E_3}|M|^2 F(E_1,E_2,E_3)(2\pi)^4\delta^4(k_1-k_2-k_3)
\end{equation}
Here indices 1, 2, and 3 refer to $n$, $\chi$ and $\phi$.  The Pauli blocking factor $F = f_1(1-f_2) - f_2(1-f_1)$ considering both forward and reverse process, where $f_i = 1/(e^{(E_i-\mu_i)/T}+1)$. $|M|^2$ is the spin averaged matrix element squared for the process~\ref{eqn:darkdecay}. However this matrix element for the dark process is unknown. For instance, in case of a generic weak decay of the form $M = \Bar{u}(k_2)(A+B\gamma^5)u(k_1)$ \cite{KaonBV}, one cannot determine the equilibration rate as the coefficients $A$ and $B$ would be unknown. 
It may be noted that at the tree level, the matrix element 
 adds no explicit temperature dependence ($A$ and $B$ are constants), but the main contribution comes from the phase space. For the weak process $n \to p + K^-$ , the rate was found to be independent of temperature \cite{KaonBV}. 

Therefore in this scenario, we consider a constant timescale for the neutron decay. We take the rate of neutron decay to DM to be 100 times smaller than the neutron decay rate calculated from the Standard Model~\cite{Fornal2019,Chang_2018}, irrespective of the temperature inside the NS:
\begin{equation}\label{eq:tau_const}
    \tau_{n\rightarrow \chi} = \tau_{n\rightarrow p}^{SM}\times 100 = 8.8\times 10^4 s~.
\end{equation}
With this constant timescale assumption, we calculated that bulk viscosity for this decay channel, shown in Fig.~\ref{fig:Strength}. One can see that the bulk viscosity in this scenario (green curve) dominates over the m-URCA bulk viscosity for both nucleonic (blue curve) and dark (orange curve) sectors, in the temperature range of  $\sim 10^8- 10^{10}$K. \\

{\it Scenario 3}: In nucleonic matter, d-URCA is allowed only beyond a certain threshold condition satisfied for very massive NSs, and therefore m-URCA process is the dominant weak decay channel. In this scenario, we consider the case when the rate for the neutron decay via the dark channel is 1/100 times the rate of neutron decay via d-URCA process. Then one can write down the relaxation timescale as:
\begin{equation}\label{tau_dUrca}
    \frac{1}{\tau_{n\rightarrow \chi}} = \frac{\Delta\Gamma_{n\rightarrow \chi}}{ \delta \mu} \frac{\delta \mu}{n_B\delta x_n} = \frac{\Delta\Gamma_{d-URCA}/100}{ \delta \mu} \frac{\delta \mu}{n_B\delta x_n}~.
\end{equation}
From the  bulk viscosity calculation for the direct URCA process~\cite{Reisenegger_1995}, up to leading order in $\delta \mu$, we have
\begin{equation}\label{Reisenegger}
    \frac{\Delta\Gamma_{d-URCA}}{\delta \mu} = \lambda_{d-URCA} = \lambda_{m-URCA} \times 1.5\times 10^8/T_8^2
\end{equation}

where $T_8 = T/10^8 K$ and $\lambda_{m-URCA}$ is given in the above Eqn.~\eqref{Sawyer}. In Fig.~\ref{fig:Strength}, we can see that the bulk viscosity in this scenario (red curve) always dominates over the m-URCA bulk viscosity for both nucleonic and DM sectors in the temperature range of  $\sim 10^8- 10^{11}$K considered.
\\

%This paper considers two scenarios for the neutron dark decay timescale: 1) 100 times the neutron beta decay timescale \cite{Fornal2019} and 2) 100 times the m-URCA timescale. 
For the scenarios considered above, we examine whether they are in conflict with the constraints from observation of the neutrino pulse in the supernova event SN1987A~\cite{RaffeltSeckel1988}, which lasted $\sim 10-12 s$.
%The supernova SN1987A is the first observation of a neutrino pulse from a collapsing star. The neutrino pulse from the SN lasted for around 10 seconds, with half the energy lost inthe first second itself \cite{RaffeltSeckel1988}.
In scenarios 1 and 3 where the reaction rate is 100 times less than that of the leptonic process, the cooling via the exotic channel $\phi$ would be 100 times weaker than cooling via neutrinos. In the second scenario (Eq. \ref{eq:tau_const}), the neutron decay timescale $\tau_{n\rightarrow \chi} \sim 10^5$ s. If we allow the cooling via this channel, the timescale is too large to affect the cooling dynamics of the SN1987A remnant. 
Therefore SN1987A observations do not impose any constraints on the cooling considered in our work.

\subsection{R-mode instability window in presence of DM}
$R$-mode is a toroidal fluid oscillation mode in NSs that may become unstable due to the Chandrasekhar-Friedman-Schutz (CFS) mechanism~\cite{CFS1,CFS2} leading to continuous GW emission. Depending on timescales, viscous processes (shear and bulk viscosity) of  NS  matter may result in  damping  these  oscillation  modes. The balance between viscous and gravitational timescales defines an instability window within which the $r$-mode amplitude can grow, spinning down the star~\cite{owen1998gravitational}. In our model of hadronic matter, we consider only the shear viscosity due to the n-n scattering and bulk viscosity due to m-URCA reactions although several other sources of viscosity e.g. non-leptonic weak interactions involving hyperons~\cite{Lindblom2002,Debi_2006}, ungapped interacting quark matter~\cite{Alford2014}, Ekman layer damping~\cite{ShterninYakovlev2008} can affect the $r$-mode instability window. In addition, we consider contributions to shear viscosity from DM self-interaction and bulk viscosity from neutron decay as damping source to the $r$-mode instability. The instability window boundary is determined by conservation equations that equate the energy loss rates, namely, the power $P_G$ fed into the $r$-mode by radiating gravitational waves and the dissipated power $P_D$ from the viscous mechanisms~\cite{owen1998gravitational,Alford2014}
\begin{equation}
    P_G = P_D \rvert_{\alpha \rightarrow 0}
\end{equation}
where $\alpha$ is the $r$-mode amplitude.
\\

\begin{figure}[ht]
    \centering
    \includegraphics[scale=0.8]{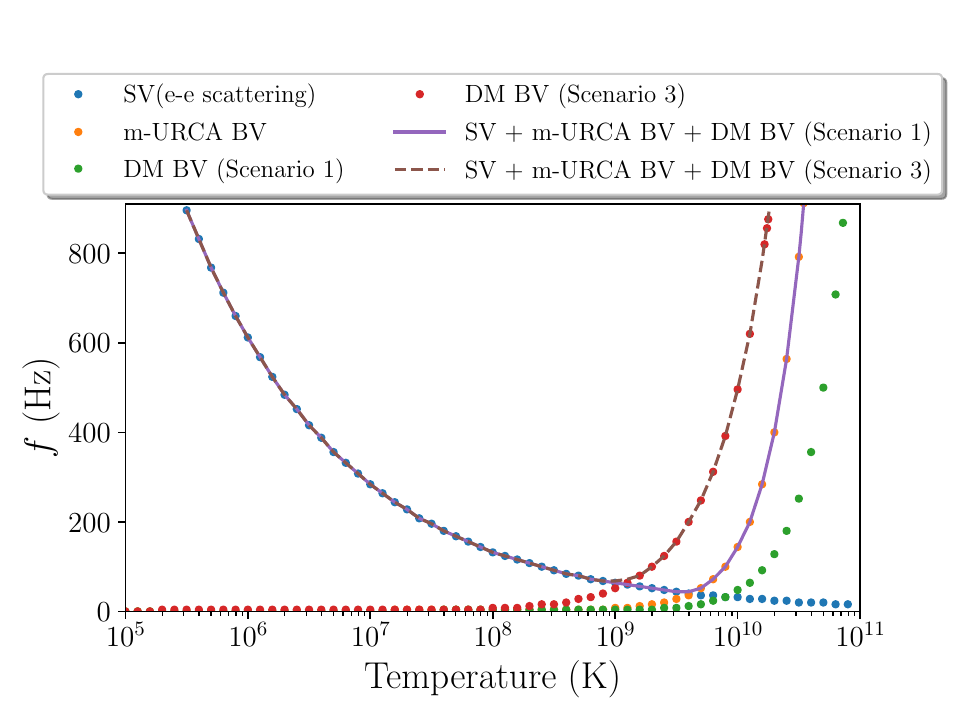}
    \caption{$R$-mode static instability window in the $f-T$ plane for 1.4M$_{\odot}$ NS for the different scenarios of viscous dissipation inside the NS. The dotted points with different colours show individual contributions and the solid and  dashed lines show their combined contribution for the scenarios 1 and 3 respectively as described in Sec.~\ref{sec:bulk}.
    % \js{it is hard to distinguish between the different lines and colors as they are on top of each other!}\sss{done}
    }
    \label{fig:dm_bv_all}
\end{figure}

To investigate the effect of DM on the $r$-mode instability window systematically, we vary the parameters of the DM model, namely the self-interaction strength $G$ and the damping timescale ($\tau_{n\rightarrow \chi}$ denoted by $\tau$ from here on). We found that the parameter $G$ has a negligible effect on the $r$-mode instability boundary (figure not shown). For the calculation of the neutron decay timescale, we consider the three different scenarios described in Sec.~\ref{sec:bulk}. In Fig.~\ref{fig:dm_bv_all}, we plot the $r$-mode instability window in the $f-T$ (frequency vs temperature) parameter space comparing the effect of individual viscosity sources in the Fig.~\ref{fig:Strength}.  The curves represent a $1.4M_{\odot}$ NS with HTZCS parametrization for the hadronic EOS and $G$ = 26 fm$^2$ self-interaction strength for the DM EOS.
\\

\begin{figure}[ht]
    \centering
    \includegraphics[scale=0.8]{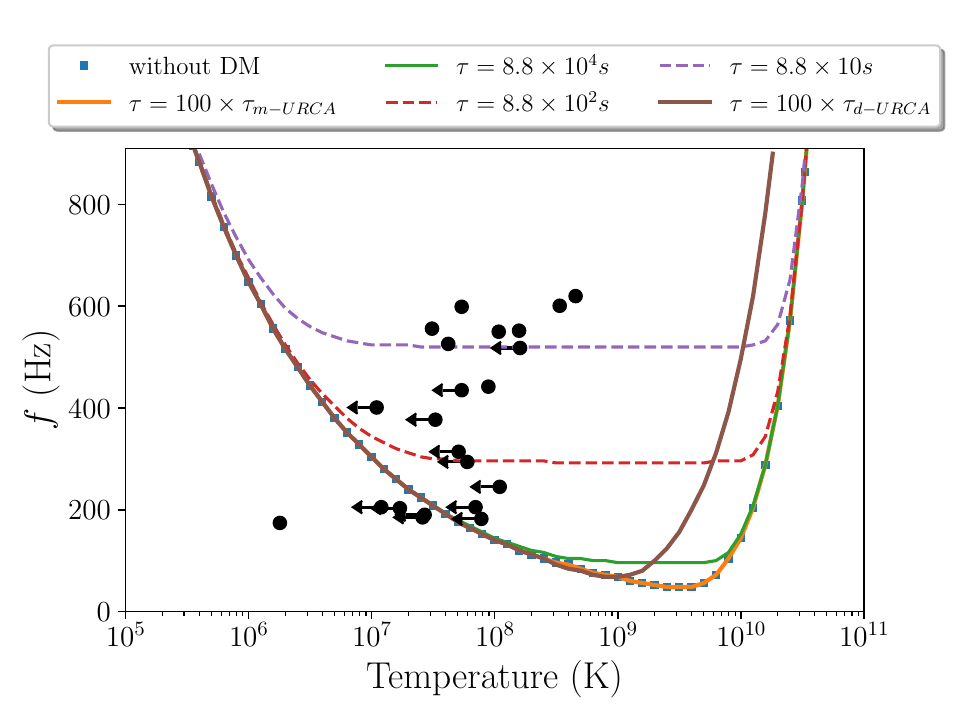}
    \caption{$R$-mode static instability window in the $f-T$ plane for 1.4M$_{\odot}$ NS for the three different scenarios of relaxation timescale $\tau$ described in Sec.~\ref{sec:bulk} along with the X-ray data without the error estimates~\cite{Haskell2012}. 
    % \js{note that the two cases ... are on top of each other.} \sss{done}
    }
    \label{fig:dm_bv_fT}
\end{figure}

In nucleonic matter, shear viscosity from e-e scattering (blue dotted line) and bulk viscosity from m-URCA reactions (orange dotted line) are considered. For DM bulk viscosity, two scenarios 1 and 3 are considered where the rate is 1/100 times of nucleonic m-URCA (green dotted line) and d-URCA (red dotted line) reactions, respectively.
The total instability window is defined by the solid and dashed lines, bounded by the shear viscosity at low temperatures and bulk viscosity at high temperatures.
In scenario 1, the nucleonic bulk viscosity dominates over the DM bulk viscosity, and therefore the total instability window (purple solid line) coincides with the minimal hadronic scenario.
On the other hand, in scenario 3, DM bulk viscosity dominates over nucleonic m-URCA bulk viscosity, resulting in shrinking of the total instability window (brown dashed line) as compared to the hadronic case.
%It is evident that for the first scenario ($\tau = 100\times \tau_{m-URCA}$), there is no visible change in the instability window because the DM viscosities (both shear and bulk) never dominate the nucleonic contributions as also shown in the Fig.~\ref{fig:Strength}. But for the third scenario ($\tau = 100\times \tau_{d-URCA}$), we see that the boundary shrinks in the low temperature region where the DM bulk viscosity dominates (refer to Fig.~\ref{fig:Strength}).
In addition to the scenarios 1 (orange solid line) and 3 (brown solid line) shown in Fig.~\ref{fig:dm_bv_fT}, we also plot scenario 2 (constant timescale). Since the timescale in this scenario is independent of temperature, we get a constant frequency boundary in this region. To demonstrate the sensitivity of the instability window to the value of $\tau$, we consider three different values of $\tau = 8.8\times 10 s$ (purple dashed line), $8.8\times 10^2 s$ (red dashed line) and $8.8\times 10^4 s$ (green solid line).
\\
%to demonstrate that in this scenario, the instability boundary remains constant when the DM BV dominates and is at a much higher frequency with decreasing $\tau$. 

% \sss{\sout{In Fig.~\ref{fig:limit}, we also plot the minimum frequency of the instability boundary as a function of the decay timescale for the constant timescale scenario for bulk viscosity and see that the frequency boundary value decreases with increasing $\tau$. At the expected decay rate of $\tau_{n\rightarrow \chi} = \tau_{n\rightarrow p}^{SM}\times 100 $ from Eqn.~\eqref{eq:tau_const}, we see the boundary frequency value is around $0.1\times \Omega_K$ which is very close to the non-DM scenario.}} 

\begin{figure}[ht]
    \centering
    \includegraphics[scale=0.8]{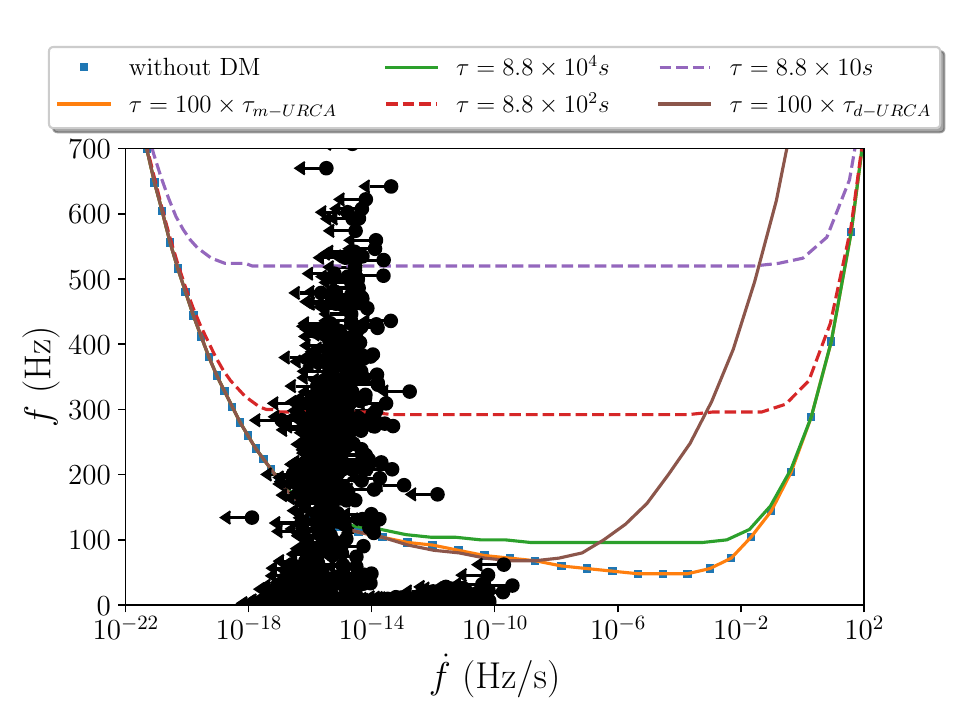}
    \caption{$R$-mode dynamic instability window in the $f-\Dot{f}$ plane for a 1.4M$_{\odot}$ NS for the three different calculations of the relaxation timescale $\tau$ as described in Sec.~\ref{sec:bulk} along with the pulsar timing data from the ATNF catalogue~\cite{ATNF}}
    \label{fig:dm_bv_ff1}
\end{figure}

To check whether the instability windows considered are consistent with recent pulsar observations, in Fig.~\ref{fig:dm_bv_fT} we plot the $f-T$ data (without the error bars as estimated from the envelope model) for low-mass x-ray binaries (LMXBs)~\cite{Alford2014,Haskell2012}, that are heated and potentially spun up by accretion from a companion. Further, to take into account pulsar timing data, we convert this $f-T$ boundary to a dynamical boundary in $f-\Dot{f}$ parameter space by assuming that the power-loss due to the spin-down driven by $r$-mode instability is equal to the luminosity (both neutrino and photon luminosity) of the star
\begin{equation}
    P_G = L_{\nu} + L_{photon}
\end{equation}
where $L_{\nu}$ and $L_{photon}$ denote the neutrino and photon luminosity~\cite{luminosity,Alford2014a}, respectively.
In Fig.~\ref{fig:dm_bv_ff1}, we plot the same instability window in the $f-\Dot{f}$ plane along with the pulsar timing data from the ATNF catalog~\cite{ATNF}.  Since we assumed that the spin down is entirely due to $r$-modes ignoring other spin-down mechanisms, we put a left arrow in the $\Dot{f}$ parameter for all the pulsars indicating that the spin-down $\Dot{f}$ originating from $r$-modes alone will be less than that measured in pulsar data. For both the Figs.~\ref{fig:dm_bv_fT} and \ref{fig:dm_bv_ff1}, we see that the $r$-mode instability cannot be reconciled with the pulsar data for currently known minimal hadronic damping mechanisms. 
Adding DM contribution, the instability window in scenarios 1 and 3 fail to explain recent observational data. However, in scenario 2 (constant timescale), low values of $\tau$ is favoured by pulsar data. Although among the values considered, only $\tau = 8.8\times 10^4 s$ is consistent with experimental data from neutron decay anomaly, lower $\tau$ values may be possible considering other decay channels.

%%%%%%%%%%%%%%%%%%%%%%%%%%%%%%%%%%%%%%%%%%%%%%%%%%%%%%%%%%
\section{Discussions}
\label{sec:discussions}
In this work, we perform the first systematic investigation of the effect of dark matter in NSs on $r$-mode instability. For the hadronic model, we consider state-of-the-art equations of state that are compatible with constraints from chiral EFT calculations and multi-messenger (electromagnetic and gravitational wave) observations. The shear and bulk viscosity in hadronic matter taken into account comes mainly from leptonic weak interaction processes and is well known. 
For the DM model, we adopt an EOS motivated by the neutron decay anomaly explained via a dark decay channel. The decay rate of neutrons via the dark channel fixes the key ingredient for the calculation of bulk viscosity in presence of DM. The shear viscosity in the presence of DM can be estimated from the DM self-interaction cross section employing kinetic theory. 
Using the shear and bulk viscosities, one can determine the effect of the presence of DM on the $r$-mode instability window in NSs.

For the DM model considered, even for the most optimistic value of self-interaction strength $G$, we found that the shear viscosity of DM is orders of magnitude lower than that of hadronic matter for temperatures below $10^9$ K. For the temperature regime in which DM shear viscosity becomes important (T > $10^{10}$ K) damping is already dominated by hadronic bulk viscosity. Therefore, we can conclude from our investigation that the effect of DM shear viscosity can be neglected for the model considered.

For the bulk viscosity calculation for DM in NSs,
we compared the strength of the DM bulk viscosity against the modified URCA bulk viscosity at different temperatures for each scenario.
% and found that DM bulk viscosity is always several orders of magnitude lower than that of the m-URCA bulk visosity. 
For the calculation of the $r$-mode instability window, we investigate the role of both the self-interaction strength $G$ and the relaxation timescale $\tau$. We found that the parameter $G$ has a negligible effect on the $r$-mode instability boundary. 
 For the relaxation timescale for the neutron decay, we considered three different scenarios. In the first scenario, we assumed that the relaxation rate is always 100 times smaller than that of the m-URCA reaction and follows the same temperature dependence. The alternative assumption (scenario 2) considered the neutron decay rate to protons, and the neutron decay to DM rate to be 100 times smaller than
that, irrespective of the temperature inside the NS. This assumption leads to a constant
timescale for the neutron decay. 
It is interesting to see how the BV and instability window respond to the different timescales. As the temperature dependence of the timescale is unknown, this assumption is particularly helpful to get an idea of how the BV would be modified for arbitrary temperature dependence. It demonstrates that the instability window can be drastically modified. In the third scenario, the DM relaxation rate follows the temperature dependence of d-URCA reaction.
%and the data becomes consistent only if the timescale is low (see Figs. \ref{fig:dm_bv_fT}, \ref{fig:limit}) i.e. rate is fast.

For all the three scenarios, we compared the instability windows with and without the presence of DM against the x-ray data and pulsar data from the ATNF catalogue. We concluded that the instability window with minimal hadronic damping mechanisms on the inclusion of shear and bulk viscosity contributions from DM within scenarios 1 and 3, remained incompatible with the observational data. However, we observed that if the dark decay is much faster than the neutron decay timescale in scenario 2, the instability window gets significantly reduced and could explain the observed pulsar data. This could be possible for alternative non-$\phi$ reaction channels (e.g. $n+\chi \to \chi + \chi)$, similar to non-mesonic decay channels in hyperons ($\Lambda + N \to N + N$), but we leave this for a future study.

While there are many studies in the literature calculating shear and bulk viscosity of hadronic and quark matter, there are to our knowledge no prior studies of DM viscosities on $r$-modes. In this work, the choice of the DM model was motivated by the recent work of~\cite{motta2018a, motta2018b, husain2022a}, which were based on the neutron decay anomaly and considered DM self-interaction via a $U(1)$ vector gauge boson. This work used the Quark Meson Coupling model for the hadronic EOS and imposed constraints from the maximum NS mass to deduce an upper limit of 26 fm$^2$ for the value of $G$. In our case, we considered a hadronic model compatible with recent  chiral EFT and multi-messenger constraints and derive a lower limit of 5.6 fm$^2$ for the value of $G$. There is  just a couple of earlier work~\cite{horowitz2012, yoshida2020} that derived the relation between the DM shear viscosity and the DM cross-section using kinetic theory. In our calculations, we used the derived value of lower limit for $G$ to obtain an estimate of the DM shear viscosity.

The results of this investigation are timely and interesting for both the observational pulsar astrophysics and gravitational wave communities. There is a considerable  interest about the effect of the presence of DM in NSs on gravitational wave signals, and $r$-modes are targets for current searches as sources of continuous GWs \cite{fesik2020, fesik2020erratum, abbott2021, rajbhandari2021}.
% \js{can you cite a paper from LIGO on that, please?} \sss{done}. 
However, calculations of DM transport properties are challenging given that there are a large number of DM models spanning wide ranges of properties of DM candidates, given that current observational data from astrophysics and cosmology do not have a high constraining power. 
% \sss{This is a concept paper demonstrating the evaluation of BV and $r$-mode instability window for a particular DM model and for modified and direct URCA temperature dependence of the timescale. We can further have different interactions describing this microscopic dark decay process.} 
While this study lays the foundation for calculating the viscosity of DM in NSs for particular chosen models, this also opens up the possibility of investigating other hadronic and DM models. 
In particular, the inclusion of an attractive two-body interaction between the DM particles will form Cooper pairs of DM particles which could be able to describe the pulsar data by including effects from a dark superfluid phase. We plan to investigate this possibility in forthcoming work.

\acknowledgments

% \textcolor{purple}{JS: We thank the anonymous referee for pointing out ...} 
We thank the anonymous referee for the meticulous reading of our manuscript and the useful suggestions,
%and pointing out the possibility of studying scenario 3 of the direct URCA temperature dependence,
which significantly helped to improve the manuscript.
The authors thank Dhruv Pathak for providing the pulsar X-ray and timing data. The authors also thank Robin Diedrichs and Edwin Genoud-Prachex for useful discussions during this work. L.S., D.C., S.S. and S.G. acknowledge the support by the State of Hesse within the Research Cluster ELEMENTS (Project ID 500/10.006).
J.S.B. acknowledges support by the Deutsche Forschungsgemeinschaft (DFG, German Research Foundation) through the CRC-TR 211 `Strong-interaction matter under extreme conditions' -- project number 315477589 -- TRR 211. 
D.C., S.S., S.G. acknowledge usage of the IUCAA HPC computing facility for the numerical calculations. They also thank the Institute for Theoretical Physics at the Goethe University and the Frankfurt Institute for Advanced Studies in Frankfurt, Germany, for their warm hospitality, where part of this research was carried out.


% \bibliographystyle{unsrtnat}
\bibliographystyle{JHEP}
\bibliography{Refs}

\iffalse

\fi

\end{document}